\documentclass[aip, amsmath,amssymb,reprint,graphicx]{revtex4-1}
\usepackage[utf8]{inputenc}
\usepackage[T1]{fontenc}
\usepackage{mathptmx}
\usepackage{hyperref}
\usepackage{xcolor}
\usepackage{graphicx}% Include figure files
\usepackage{dcolumn}% Align table columns on decimal point
\usepackage{bm}% bold math
\usepackage{physics}

\begin{document}

\title{Cavity magnon polariton based precision magnetometry}

\author{N.~Crescini}
	\email{nicolo.crescini@phd.unipd.it}
	\affiliation{INFN-LNL, Viale dell'Universit\`a 2, 35020 Legnaro (PD), Italy}
	\affiliation{Dipartimento di Fisica e Astronomia, Via Marzolo 8, 35131 Padova, Italy}
\author{C.~Braggio}
	\affiliation{Dipartimento di Fisica e Astronomia, Via Marzolo 8, 35131 Padova, Italy}
	\affiliation{INFN-Sezione di Padova, Via Marzolo 8, 35131 Padova, Italy}
\author{G.~Carugno}
	\affiliation{Dipartimento di Fisica e Astronomia, Via Marzolo 8, 35131 Padova, Italy}
	\affiliation{INFN-Sezione di Padova, Via Marzolo 8, 35131 Padova, Italy}
\author{A.~Ortolan}
	\affiliation{INFN-LNL, Viale dell'Universit\`a 2, 35020 Legnaro (PD), Italy}
\author{G.~Ruoso}
	\affiliation{INFN-LNL, Viale dell'Universit\`a 2, 35020 Legnaro (PD), Italy}

\date{\today}

\begin{abstract}
A photon-magnon hybrid system can be realized by coupling the electron spin resonance of a magnetic material to a microwave cavity mode. The quasiparticles associated with the system dynamics are the cavity magnon polaritons, which arise from the mixing of strongly coupled magnons and photons. We illustrate how these particles can be used to probe the magnetization of a sample with a remarkable sensitivity, devising suitable spin-magnetometers which ultimately can be used to directly assess oscillating magnetic fields. Specifically, the capability of cavity magnon polaritons of converting magnetic excitations to electromagnetic ones, allows for translating to magnetism the quantum-limited sensitivity reached by state-of-the-art microwave detectors.
Here we employ hybrid systems composed of microwave cavities and ferrimagnetic spheres, to experimentally implement two types of novel spin-magnetometers.
\end{abstract}

\maketitle

Among the most studied types of hybrid systems, an important role is played by photon-magnon hybrid systems (PMHSs)\cite{Clerk2020,Lachance_Quirion_2019}. These yielded remarkable results in the study of light-matter interaction\cite{haroche}, and in the last decades emerged as promising constituents for new quantum technologies as well\cite{Chumak2015,RevModPhys.89.035002,Kurizki3866}.
PMHSs have different forms, as they are built with miscellaneous building blocks, but the underlying physics is similar.
In a magnetic field $B_0$, a spin can change its quantum state from $-1/2$ to a $+1/2$ by absorbing a spin-1 boson, like a photon, and vice versa by emitting one. In this sense, a quanta of spin excitation with energy $\hbar \omega_m = \mu_B B_0$ can be effectively described as a quasiparticle, known as magnon, which can turn into a photon of the same energy $\hbar \omega_c$\cite{PhysRev.143.372}. This reciprocal conversion is quantified by the interaction strength $g_{cm}$, known as vacuum Rabi splitting, which is the rate at which magnons are converted into photons and vice versa.
When $g_{cm}$ is much larger than the damping rates of the magnon $\gamma_m$ and of the photon $\gamma_c$, the system is in the strong coupling regime, and the quasiparticles arising from this mixing are known as cavity magnon polaritons (CMPs)\cite{Kittel2004,walls2007quantum}.

PMHSs are widely investigated for advancing quantum information science. In this field their importance lies in building quantum memories\cite{TABUCHI2016729,Zhang2015b,PhysRevLett.105.140501,bonizzoni,PhysRevLett.104.077202,bonizzoni2,doi:10.1080/09500340.2016.1148212}, in converting microwaves to optical photons\cite{Kimble2008,PhysRevLett.113.203601,PhysRevA.92.062313,PhysRevB.93.174427,cate}, or in quantum sensing, where the detection of single magnons was recently demonstrated\cite{Lachance-Quirione1603150,Lachance-Quirion425,wolski2020dissipationbased}.
CMP recently found new applications in the field of non-Hermitian physics\cite{Bender_2007,PhysRevLett.80.5243,Ruter2010}, where they already yielded outstanding results\cite{Zhang2017}. Exceptional points, spots of the system's parameter space highly sensitive to external stimulations, can be probed with PMHSs\cite{PhysRevLett.123.237202,PhysRevX.6.021007}, and new configurations may be designed to access more exotic phenomena and study their applications\cite{PhysRevB.99.054404,PhysRevB.99.214415}.
The potential of hybrid systems was also shown in many other applications of quantum physics\cite{Rao2019,PhysRevA.93.021803,PhysRevLett.123.127202,PhysRevLett.120.057202,PhysRevLett.124.053602}. 

A distinguished physical realisation of this model can be obtained by hybridising the microwave photons of a resonant cavity with the magnons of a ferrimagnetic insulator\cite{PhysRevLett.111.127003,PhysRevLett.113.083603,PhysRevLett.113.156401,PhysRevApplied.2.054002,Zhang2015}.
Such scheme was implemented with multiple purposes, for example to develop new quantum technologies with qubits\cite{Tabuchi405,TABUCHI2016729}, or for microwave-to-optical photon conversion\cite{PhysRevB.93.174427,cate}, making it an established platform of hybrid magnonics.

\begin{figure}
	\includegraphics[width=.5\textwidth]{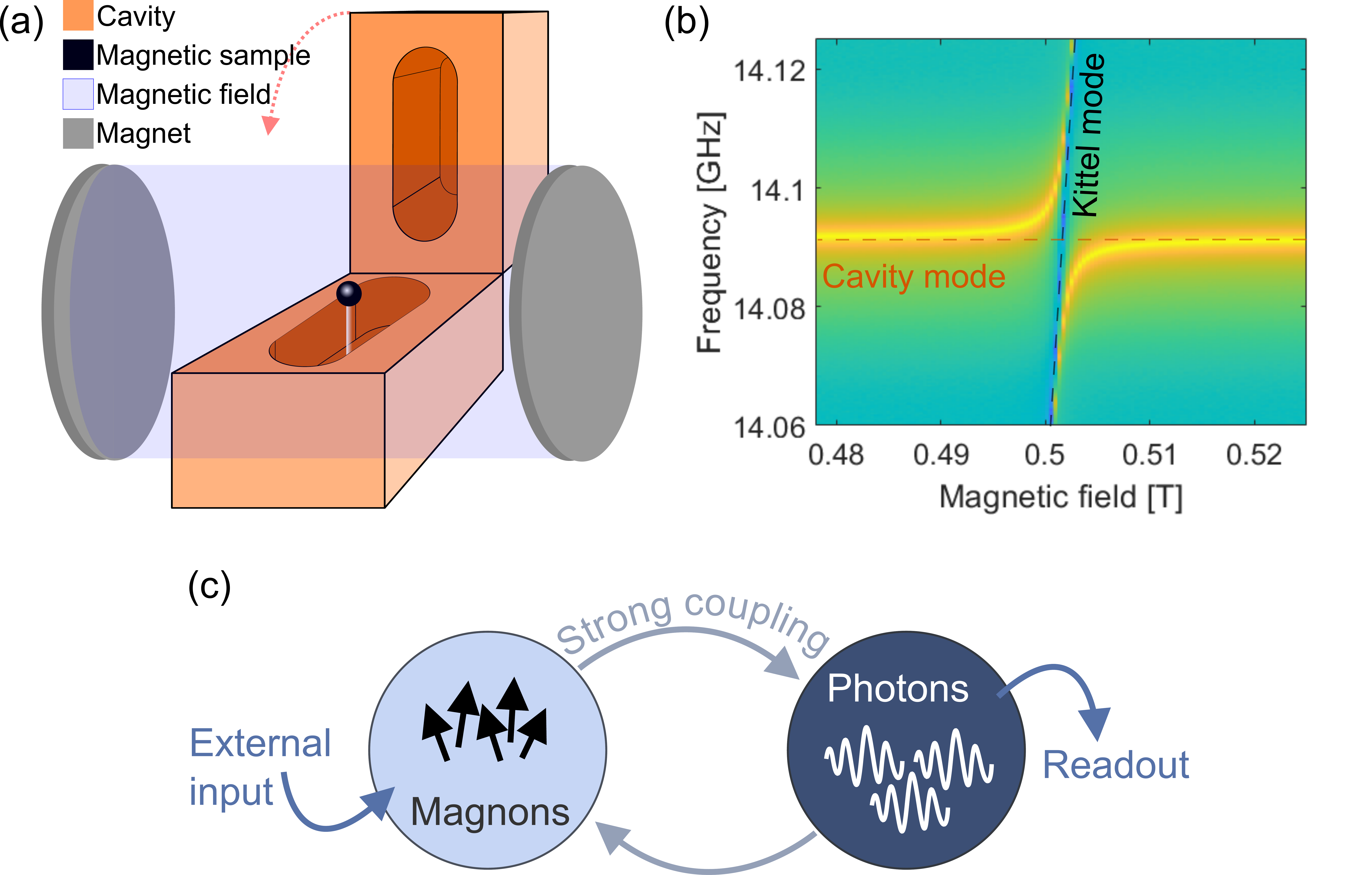}
	\caption{Schematic representation of a typical PMHS (a), anticrossing curve (b), and diagram of a spin-magnetometer working principle (c). Part (a) represents a PMHS consisting in a YIG sphere housed in a microwave cavity under a static magnetic field. Plot (b) is measured with a .5\,mm-diameter YIG sphere in a 14\,GHz copper cavity, the colour scale is in logarithmic arbitrary units, where blue to yellow is low to high transmission, and the dashed lines show the uncoupled cavity and Kittel modes.}
	\label{fig:hs}
	\end{figure}

In the devices described in this letter, we employ copper cavities as a photonic resonator and Yttrium Iron Garnet (YIG) spheres as magnetic material (see Fig.\,\ref{fig:hs}a). YIG has the exceptionally high electron spin density of $2\times10^{28}\,\mathrm{spin/m^3}$ already at room temperature, and a linewidth as narrow as 1\,MHz. This latter value is matched to the one of a typical copper cavity and, thanks to the chosen spherical shape, is not affected by geometric demagnetization.
Being employed in a number of microwave and rf devices, YIG is among the most well-known ferrites, and hence is readily available.
The magnetic sample is placed inside the cavity, where the rf magnetic field is maximum for the selected cavity mode, and is magnetised with a static field $B_0$ perpendicular to the cavity one.
In this way, the Kittel mode of magnetisation couples to the microwave cavity photons, and the system exhibits the typical anticrossing dispersion relation, of which an example is shown in Fig.\,\ref{fig:hs}b. The coupling strength depends on the working frequency, on the microwave mode volume, and on the number of spins involved\cite{PhysRevLett.113.156401}, but it is normally large enough to let the photon (magnon) oscillate into magnon (photon) many times before being dissipated.

%\textcolor{red}{
This feature of CMP to be a mixed state of microwaves and spin excitations allows one to extract information on magnons by monitoring photons.  In the presence of a strong coupling, the signal transduction is efficient, i.\,e. without signal loss, as a spin excitation is more likely converted to a photon and detected, than it is to be dissipated due to the PMHS losses (see Fig.\,\ref{fig:hs}c for a schematic diagram).
%}
Amongst other techniques to measure spin-waves\cite{Chumak2015}, the use of CMP is a particularly simple approach which exploits the sensitivity of microwave technology and transfers it to the detection of magnons.
The strong coupling makes the energy stored in a cavity dependent on the one in the material, so an antenna coupled to the electromagnetic field of the cavity gives a simple access to the features of the spin system\cite{Wolz2020}.
Nowadays electronics is extremely developed, and the detection of electromagnetic radiation has been brought to the standard quantum limit of linear amplifiers. At microwave frequencies, Josephson Parametric Amplifiers (JPA) were demonstrated to be the best devices to measure tiniest amounts of power\cite{ROY2016740}.
%\textcolor{red}{
Thanks to CMPs, such precision can be shifted to a magnetic measurement, as the electromagnetic power in the cavity is highly dependent on the magnetisation of the sample when the coupling strength largely exceeds the system dissipations $g_{cm}\gg \gamma_m,\gamma_c$. It follows that, under these conditions, the quantum-limited readout of a JPA can be exploited to detect spin excitations.
%}
	
At microwave frequencies, measuring a sample's magnetization becomes increasingly difficult because of technological limitations and fundamental problems, like for example radiation damping\cite{doi:10.1063/1.1722859,AUGUSTINE2002111,PhysRev.95.8}. In free space, radiation damping consists in the magnetic dipole emission of a magnetised sample which, at GHz frequencies, drastically decreases the coherence time, limiting the experimental sensitivity. This effect is avoided in PMHSs, as the sample is housed in a resonant cavity which removes the damping by inhibiting the phase space of the emission\cite{BARBIERI2017135}.

For all their characteristics, PMHSs emerge as an outstanding platform for precision magnetic measurement, which are of interest for a broad range of applications as well as for approaching fundamental physics issues. 
Hereafter, we describe two types of spin-magnetometers which can be designed with hybrid systems, detail their design and report on their operation. 
%\textcolor{red}{
We notice that a high occupation number of the modes permits to treat them as classical oscillators, which is often the case throughout this work, so we rely on a classical treatment of the fields.
%}
These devices are originally meant to measure tiniest oscillation of a sample's magnetization, related for example to a Dark Matter Axion field\cite{BARBIERI1989357,BARBIERI2017135}, but can be used to assess many other physical phenomena.

\textit{Transverse spin-magnetometer (TSM). - }
Let's now focus on a hybrid system like the one of in Fig.\,\ref{fig:hs}, where a magnetised YIG sphere is placed in a microwave cavity. 
If an oscillating electromagnetic, or pseudo-electromagnetic, field $\mathbf{b}_1$ is oriented perpendicularly to the static field, its quanta can be absorbed by the hybrid magnetic mode.
As the magnetization vector $\mathbf{M}$ precesses over the static field, an excitation lying on the precession plane can resonantly interact with it, and the system evolves according to Bloch equations
\begin{equation}
\dv{\mathbf{M}}{t} = \gamma (\mathbf{M}\times\mathbf{b}_1)_\bot + \frac{\mathbf{M}}{T_s},
\label{eq:pin}
\end{equation}
where $\gamma=(2\pi) 28\,\mathrm{GHz/T}$ is the electron gyromagnetic ratio, and $T_s$ is the system relaxation time.
The driven magnetization resulting from Eq.\,(\ref{eq:pin}) is
\begin{equation}
M(t) = \gamma \mu_B n_s T_s \cos(\omega_1 t),
\label{eq:tsm_mag}
\end{equation}
where $\omega_1$ is the frequency of $\mathbf{b}_1$.
In a steady state, the power of $\mathbf{b}_1$ is absorbed, re-emitted by the magnetization, and rapidly converted into photons thanks to the strong coupling.

The optimal experimental condition is an antenna critically coupled to the cavity, which in steady state can extract up to half of the power deposited by the external field, resulting in
\begin{equation}
P_1 = \gamma \mu_B N_s \omega_1 b_1^2 T_s,
\label{eq:pout}
\end{equation}
where $N_s$ is the number of spins of the hybrid system, and the field frequency $\omega_1$ is on resonance with one of the hybrid modes.
%\textcolor{red}{
To calculate the magnetic sensitivity of the TSM, in Eq.\,(\ref{eq:pout}) we substitute the deposited power $P_1$ (in Watts) with the power sensitivity of the readout electronics $\sigma_P$ (in Watts per unit of bandwidth), and recast the equation to isolate the magnetic field. We obtain the sensitivity of the TSM
\begin{equation}
\sigma_{b_1} =\sqrt{\frac{\sigma_P}{\gamma \mu_B N_s \omega_1 T_s}},
\label{eq:b1}
\end{equation}
in Tesla per unit of bandwidth, which is the field detectable in 1 second integration time with a unitary signal-to-noise ratio.
%}
%
%\textcolor{red}{The power sensitivity $\sigma_P$ is basically due to thermodynamic fluctuations, to the amplifier additive noise, and to the quantum noise. In all the aforementioned cases, the background can be quantified as an effective noise temperature $T_n$, which is used to characterise our readout electronics.
Eq.\,(\ref{eq:b1}) also shows that the spin-magnetometer sensitivity increases for larger spin-number and longer hybrid system coherence times.
%}
This suggests the use of high quality-factor cavities and samples to get a long $T_s$, and of a large volume of high spin density magnetic material to increase $N_s$.
In this sense, we found a good compromise in YIG.
%\textcolor{red}{
The scalability of the PMHS is of fundamental importance to obtain an increased sensitivity of the setup, as it is directly related to the increment of $N_s$.
%}
To this aim we design spin-magnetometers based on multi-samples PMHS\cite{quaxepjc,PhysRevLett.124.171801}, embedded in cylindrical cavities. 
%\textcolor{red}{
To further boost the magnetic sensitivity, we reduce $\sigma_P$ by operating the device at milli-Kelvin temperatures, to reduce thermal noises and to consent the use of quantum-limited amplifiers. 
%}

\begin{figure}
\includegraphics[width=.475\textwidth]{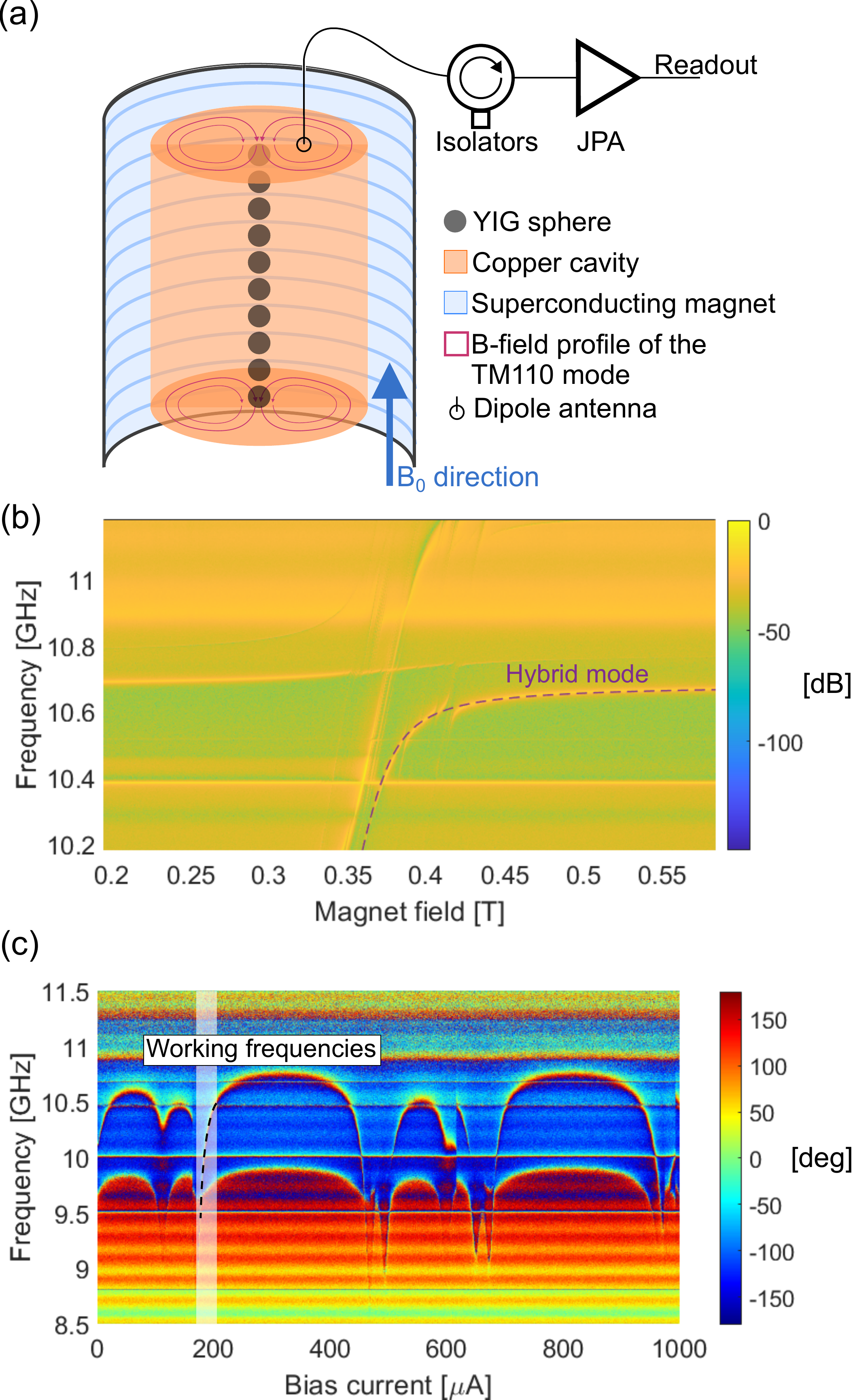}
\caption{(a) Simplified scheme of the TSM operated for an axion search\cite{PhysRevLett.124.171801} (see text for details). (b) Anticrossing curve of the 10 YIG spheres PMHS, where the dashed line indicates the low-frequency hybrid mode monitored during the measurement. (c) Phase-current diagram of the JPC mounted in this setup, here the dashed line shows the optimal working points of the amplifier. From plots (b) and (c) one notes that the 10.2-10.4\,GHz band enables both the PMHS signal transduction and the JPA amplification.}
\label{fig:tsm}
\end{figure}

Following these directions, we built a TSM whose scheme is reported in Fig.\,\ref{fig:tsm}a.
%As explained previously, in these spin-magnetometers the PMHS acts as a transducer of magnetic excitation, so we devised the largest optimally-controlled system of this kind to date\cite{crescini2020coherent}. 
%\textcolor{red}{
Its PHMS comprises ten YIG spheres, all of 2.1\,mm-diameter, produced in-house. These are biased with a magnetic field supplied by a superconducting magnet, with 7\,ppm uniformity over the volume containing the spheres. We realise the PMHS by placing the spheres along the axis of a cylindrical cavity (33\,mm-diameter, 65\,mm-length) allowing them to couple with the uniform rf magnetic field of the TM110 mode at 10.7\,GHz.
%}

%\textcolor{red}{
The PMHS has been designed to reduce the effects of the magnetic dipole interaction between different spheres and of higher order magnetostatic modes. By removing the degeneracy of the TM110 mode we limit the interference of other cavity modes; this is achieved employing a cavity with a quasi-circular section\cite{crescini2020coherent,paperdelreferee}. 
To describe this system we used a second quantisation model consisting in four coupled harmonic oscillators. We fit it to the experimental anticrossing curve of Fig.\,\ref{fig:tsm}b\cite{crescini2020coherent}.
We then operate the magnetometer in the frequency band 10.2-10.4\,GHz, part of the lower frequency hybrid mode range, as identified by the fit (dashed line in the figure)\cite{PhysRevLett.124.171801}.
The operational range is matched with the working band of our Josephson Parametric Converter (JPC), i.e. a JPA formed by a Josephson ring modulator shunted with four inductances\cite{PhysRevLett.108.147701}. 
The JPC tuning is allowed by a small superconducting coil biased with a constant current, as shown Fig.\,\ref{fig:tsm}c.
The dashed lines in Fig.\,\ref{fig:tsm}c includes the 10.2-10.4\,GHz frequency interval, showing that in this range the lower frequency hybrid mode can be monitored with our amplifier.
%}
The JPC is screened from external disturbances with different layers of superconducting and $\mu$-metal shields, and we verified that the solenoid providing the static field is not affecting the resonance frequencies of the amplifier.

%\textcolor{red}{
The noise temperature and gain of the electronics chain has been characterised with the injection of microwave signals of known amplitude in an antenna weakly coupled to the cavity. 
%The readout of calibrated signals with different amplitudes permits to quantify the noise temperature and the gain of the setup. 
The effective noise temperature results $T_n\simeq 1\,K$, which sets the noise power per unit of bandwidth $\sigma_P=k_B T_n$, where $k_B$ is Boltzmann constant. The contribution of the quantum limit to the noise budget is 0.5\,K, and the remaining 0.5\,K is consistent with extra noise added by the second-stage amplifier, by the losses of the wires and by the PMHS thermodynamic temperature of $\sim100\,$mK\cite{PhysRevLett.124.171801}.
The spin number and relaxation time are obtained by fitting our model to the transmission measurements of Fig.\,\ref{fig:tsm}b.
The measurement of $\sigma_P$, and of $N_s$ and $T_s$ through the PMHS spectroscopy, allows us to calculate the sensitivity of the TSM using Eq.\,(\ref{eq:b1}). With the parameters of this setup we obtain a magnetic sensitivity of
\begin{align}
\begin{split}
\sigma_{b_1} = & 0.9\times10^{-18} \Big[ \Big( \frac{1\,\mathrm{K}}{T_n} \Big) \Big( \frac{N_s}{10^{21}} \Big) \\
	& \Big( \frac{\omega_1/2\pi}{10.4\,\mathrm{GHz}} \Big) \Big( \frac{T_s}{168\,\mathrm{ns}} \Big) \Big]^{1/2} \mathrm{\frac{T}{\sqrt{Hz}}}.
\label{eq:tsm_sens}
\end{split}
\end{align}
%}
That the sensitivity given by Eq.\,(\ref{eq:b1}) holds if the field to be detected has two characteristics: a coherence time longer than $T_s$, and a coherence length long enough to comprise all the $N_s$ spins.

In particular, this is the case of the field induced by Dark Matter axions\cite{BARBIERI1989357,BARBIERI2017135}, which at GHz frequencies satisfies both these conditions.
%\textcolor{red}{
We used this TSM with a fixed bandwidth of 5\,kHz to search for axions, obtaining a limit on their effective field of $5.5\times10^{-19}\,\mathrm{T}$ with about ten hours of integration\cite{PhysRevLett.124.171801}.
%}
A TSM has the advantage of being sensitive to a (pseudo)magnetic field acting on a sample which is within the volume of a resonant cavity. In such a controlled environment, external electromagnetic disturbances are unlikely to be present, making it an interesting testbed for fundamental physics, which are usually not subjected to such screening.
However, from the point of view of the TSM possible technological employment, this feature is a limitation. In fact, the screening due to the cavity makes it difficult to expose the material to a field which is uniform and coherent over the magnetic material volume.
Hence, the application of this device is probably limited to the search of new physics.

\begin{figure}[h!]
\includegraphics[width=.45\textwidth]{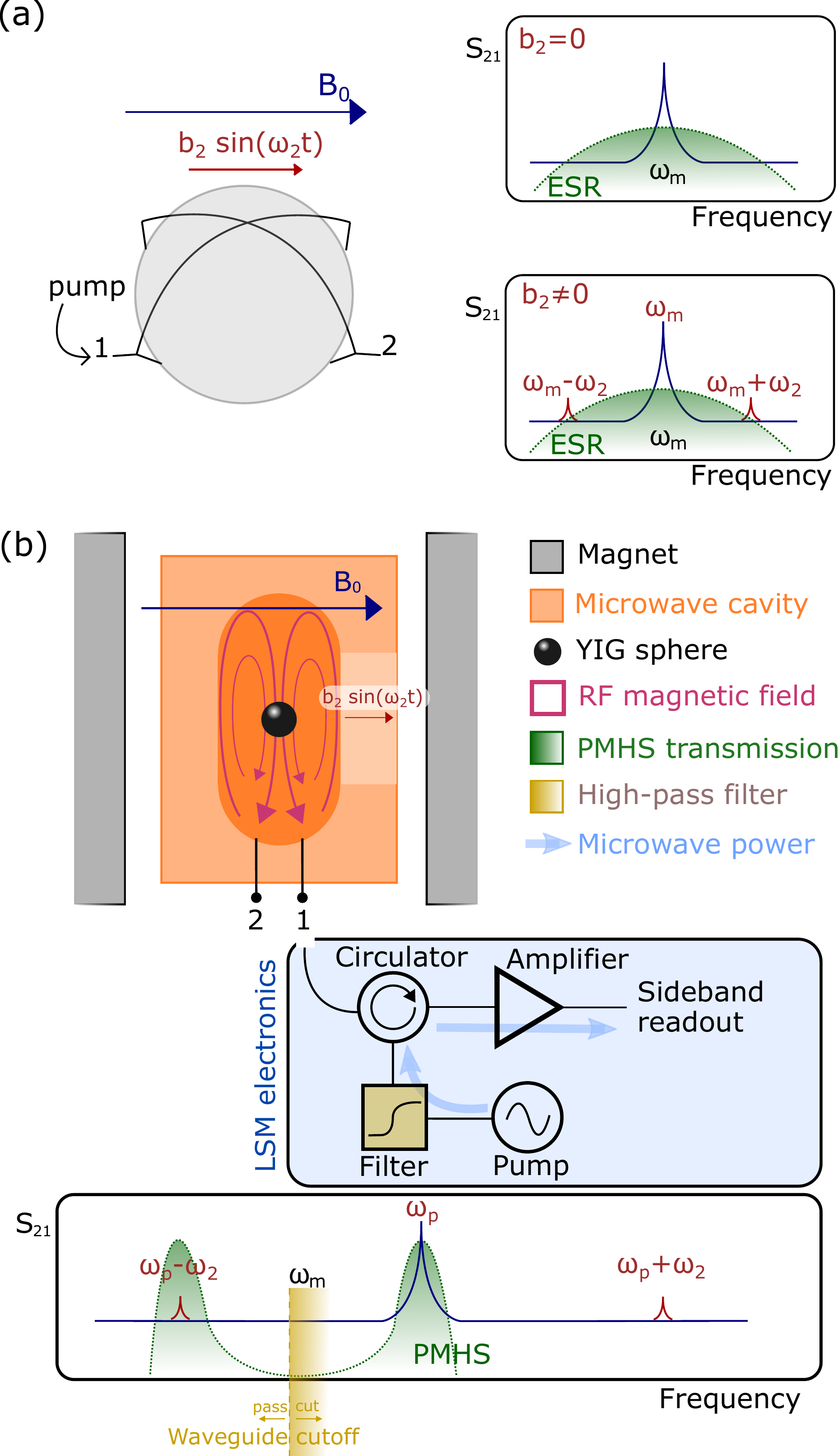}
\caption{Schematic explanation of the LSM working principles. (a) Usual design used for the detection of an ESR in a magnetic material consisting in a spherical sample (grey) surrounded by two loops in free space. The rf is fed into the system by loop antenna 1 and the output is read with the perpendicular loop 2. A pump, shown as a blue line in the spectra, is applied on-resonance with the ESR curve (green areas in the $S_{21}$ spectra).
In absence of other fields the result is a single peak (left plot), while with a superimposed oscillating field the phase of the carrier is modulated by the shifting of the ESR induced by $b_2\sin(\omega_2 t)$. Two sidebands, reported in dark red in the right plot, appear at $\omega_m\pm\omega_2$. (b) In the LSM a microwave tone is applied at the frequency of an hybrid mode, while the detection of a sideband, on-resonance with the second mode, probes the presence of $b_2$-like fields. The picture shows our room-temperature pilot setup, comprising a YIG sphere and a perforated cavity which allows for the calibration of the spin-magnetometer. Numbers 1 and 2 are two antennas coupled to the cavity, and the side of the cavity coloured in light orange represent a hole housing a loop used for calibration. See text for further details.}
\label{fig:lsm}
\end{figure}

\textit{Longitudinal spin-magnetometer (LSM). - }
In another possible measurement scheme a persistent oscillating $B$-field is parallel to the static one. In this configuration, the sample's magnetization precesses about a field $B_0+b_2\sin(\omega_2 t)$, where $\omega_2$ and $b_2$ are the oscillating field frequency and amplitude, and $t$ is time.
To illustrate the experimental arrangement, we first consider a simplified scheme including only the material and ignoring the presence of the cavity. The experimental scheme is shown in Fig.\,\ref{fig:lsm}a, where a sphere is surrounded by two crossed loops. Loop number 1 is used to excite the material, while loop number 2 senses the transmitted rf signal, and $S_{21}$ plots are measured. 
%\textcolor{red}{
The electron spin resonance (ESR) frequency $\omega_m$ of the magnetised sample is modulated at the frequency $\omega_2\ll\omega_m$ by varying the field $b_2\ll B_0$.
%}
If a monochromatic tone is applied on resonance with $\omega_m$, the effect of $b_2$ is then to transfer some of the pump power, the carrier,  to sidebands at frequencies $\omega_m\pm n \omega_2$, as schematically shown in Fig.\,\ref{fig:lsm}a for $n=1$.
In the $S_{21}$ spectrum of this simplified system, the amplitude of the first order sideband results
 %can be calculated with respect to the carrier amplitude $A_p$ and to the quality factor of the modulated resonance $Q$, as
\begin{equation}
\zeta_1 = \frac{\pi A_p^2 Q b_2}{2 B_0},
\label{eq:side}
\end{equation}
where $A_p$ is the carrier amplitude and $Q=\omega_m/\gamma_m$ the quality factor of the ESR.
%which has to be compared with the detector noise $A_n$ to obtain the magnetic field sensitivity of the device.
%Referring to Fig.\,\ref{fig:lsm}a, the first loop is used to pump the ESR, while the second one probes the sidebands with (right) or without (left) the oscillating field.
In a standard ESR technique an externally applied $b_2$ is used to detect the derivative of the ESR curve with a lock-in amplifier. Here we invert such scheme, and search for oscillating $b_2$-fields by sensing the presence of sidebands.
%\textcolor{red}{
The detection of sidebands is limited by the effective noise temperature of the system determining $\sigma_P$, the power sensitivity already defined in the case of the TSM.
The amplitude $\zeta_1$ is given by Eq.\,(\ref{eq:side}) only within the linewidth of the ESR, and drastically reduces for $\omega_2>\gamma_m$. On the other hand, when $\omega_2<\gamma_m$, extra noise induced by the pump residual amplitude modulation will increase $\sigma_P$. 
Moreover, at GHz frequencies, radiation damping broadens the linewidth of the ESR, reducing $Q$.
%}
As a consequence, in this configuration the sensitivity for measuring a $b_2$ field is poor and needs some improvements that can be engineered using PMHSs as follows.

%\textcolor{red}{
By including a cavity one may consider a PMHS's hybrid mode instead of a bare ESR.
CMPs are immune from radiation damping thus we can couple the ESR to a microwave cavity to improve the detection sensitivity.
We call $\omega_p$ one of the PMHS resonant frequencies: $\omega_p$ is also modulated by the oscillating field as $\partial \omega_p/\partial B_0 \simeq r\times\gamma$, where $0\le r\le 1$ is a field-dependent coefficient. When $\omega_m$ is equal to the cavity mode resonant frequency, $r=1/2$.
%}
The rf electromagnetic field of the PMHS, pumped with a tone on-resonance with one of the hybrid modes, i.\,e. at $\omega_p$, is phase-modulated through the variation of the resonant frequency, and therefore produces sidebands too. Their amplitude drastically decreases when they depart from the resonance frequency by several linewidths, but in a PMHS, at the frequency of the second hybrid mode, one sideband does not vanish and hence can be detected (see Fig.\,\ref{fig:lsm}b).
This device is thus sensitive to fields which are at frequencies $\omega_2\simeq 2g_{cm}$, the splitting of the two hybrid modes.
The possibility of detecting the sideband at a frequency much different from the pumping one allows us to drastically reduce the noise by heavily filtering the pump noise.
%\textcolor{red}{
A waveguide is a high pass filter which can cut low frequencies by tens of dB, and that we employ to remove the background related to the pump. 
If the sideband frequency $\omega_p-\omega_2$ is below the waveguide cut-off, its amplitude is not filtered but the pump noise is (see the electronic scheme in Fig.\,\ref{fig:lsm}b). By assuming that the carrier noise can be made lower than thermal fluctuations, the latter becomes the fundamental limitation to the apparatus sensitivity.
The magnetic sensitivity can be calculated by rephrasing Eq.\,(\ref{eq:side}), and substituting $\zeta_1$ (the sideband power) with the readout sensitivity $\sigma_P$ to obtain
\begin{equation}
\sigma_{b_2} = \frac{2B_0}{\pi r Q}\sqrt{\frac{\sigma_P}{A_p^2}},
\label{eq:lsmsens}
\end{equation}
where, in this case, $Q$ is the quality factor of the hybrid mode.
%}
From Eq.\,(\ref{eq:lsmsens}) one can see that the carrier power $A_p^2$ can be arbitrarily increased to improve the LSM magnetic sensitivity, assuming that its noise can be reduced consequently.
With realistic parameters of our PMHS, one can estimate the sensitivity of a room-temperature LSM with Eq.\,(\ref{eq:lsmsens}), resulting in
\begin{equation}
\sigma_{b_2} = 10.4\, \Big( \frac{B_0}{0.4\,\mathrm{T}} \Big) \Big( \frac{Q}{10^4} \Big) \sqrt{ \Big( \frac{A_n^2/k_B}{300\,\mathrm{K}} \Big)\Big( \frac{100\,\mathrm{mW}}{A^2_p} \Big) } \,\mathrm{\frac{fT}{\sqrt{Hz}}},
\label{eq:lsmcalc}
\end{equation}
which is already competitive with state-of-the-art magnetometers like superconducting quantum interference devices (SQUIDs) \cite{PhysRevLett.12.159,doi:10.1063/1.322574,Aprili2006,1335547,doi:10.1002/3527603646} or spin-exchange relaxation-free (SERFs) \cite{Kominis2003,PhysRevLett.95.063004,Budker2007}.
%\textcolor{red}{
Interestingly, $\sigma_{b_2}$ does not depend on any extensive parameter, in contrast with $\sigma_{b_1}$ which relies on the total number of spins. This means that the LSM can in principle be miniaturised without compromising its sensitivity, and removing the need of detecting a uniform field over a large volume.
%}

A room-temperature prototype was devised to test the actual functioning of this device, and a scheme of the setup is reported in Fig.\,\ref{fig:lsm}b. The number of spins in the sphere, together with the shape of the rf-magnetic field of the cavity mode, set the magnetometer working frequency $\omega_2\simeq(2\pi)\,200\,\mathrm{MHz}$. 
%\textcolor{red}{
The cavity mode and the ESR resonate at 11.5\,GHz (corresponding to $B_0=0.4\,\mathrm{T}$), and their linewidths determine the overall quality factor of the hybrid mode $Q=2750$, which is approximately the average of the two.
An antenna with variable coupling is connected to the cavity to inject and extract power from the hybrid system through a circulator. A microwave pump on resonance with the high-frequency hybrid mode at $\omega_p$ is filtered with a waveguide before being injected in the PMHS, obtaining the input power $A_p^2=0.2\,\mathrm{mW}$. 
The signal to be detected is the PMHS output power of the sideband at $\omega_p-\omega_2$, the lower hybrid mode frequency. At $\omega_p-\omega_2$, the background is mainly thermal thanks to the filtering waveguide.
The extracted signal is amplified with a low noise HEMT before being acquired with a spectrum analyser, and the whole electronic chain has been characterised by injecting calibrated signals.
The readout noise results $\sigma_P \simeq4\times10^{-21} \,\mathrm{W/Hz}$, mostly due to room temperature thermodynamic fluctuations, and two orders of magnitude lower than the pump noise, showing that our configuration almost removes the tone-induced background.
To calibrate the magnetic sensitivity of the device, we inject pico-Tesla fields at 200\,MHz using a single loop on one side of the cavity, which generates a known field parallel to $B_0$ on the YIG sphere (see Fig.\,\ref{fig:lsm}b).
%}
The setup was not optimized, but the expected losses due to imperfect matchings can be measured and accounted for by a factor $k=2.1$, lowering the LSM sensitivity.
With these quantities, from Eq.\,(\ref{eq:lsmsens}), the expected sensitivity of the apparatus results $k\sigma_{b_2}=1.9\,\mathrm{pT/\sqrt{Hz}}$. The prototype was calibrated with fields ranging from 2 to 14\,pT and shows a measured sensitivity of $2.0\pm0.4\,\mathrm{pT/\sqrt{Hz}}$, compatible with the estimated value\cite{inprepmag}.
In this setup the loop on the side of the cavity was used for calibration, but in principle, it can be an input coil which transduces a field to the sensitive element of the magnetometer (the magnetic sphere). This signal transduction is similar to what is usually done with SQUIDs, where an input coil is coupled to the junctions loop.
%\textcolor{red}{
From Eq.\,(\ref{eq:lsmsens}), one notices that the LSM magnetic sensitivity benefits from high quality factors, low readout noise, and high pump power. 
The former feature is related to the quality factors of the PMHS, which should comprise narrow-linewidth cavities and magnetic materials to improve the magnetometer sensitivity.
The latter two essentially depend on the microwave electronics of the setup.
Since the sensitivity is size-independent, miniaturisation can be foreseen by using 2D printed resonators and small quantities of material. Eventually, we mention that multiplexing was also shown to be a viable option in similar devices\cite{doi:10.1063/1.4973872,doi:10.1063/1.1791733,doi:10.1063/1.2803852}.
%}

%\textcolor{red}{
The sensitivity of the two PMHS-based magnetometers is limited by the noise of the readout noise temperature, which ultimately consists in quantum fluctuations\cite{PhysRevD.88.035020}. 
%}
We foresee the use of broadband Travelling Wave JPA\cite{CULLEN1958, Macklin307, PhysRevX.10.021021} to overcome the standard JPA limitation of being resonant.
%\textcolor{red}{
To overcome the quantum-limit one may rely on single photon or magnon counters, which are unaffected by this issue, rather than on linear amplifiers.
%}

%\textcolor{red}{
A downside of both the magnetometers is that their resonant nature implies a reduced bandiwdth, limited to the linewidth of an hybrid mode. Nevertheless, as the resonant frequencies of the hybrid modes can be changed with a tuning of the $B_0$ field, the band of both the TSM and LSM can be extended. 
In particular, for TSMs this changes the hybrid mode frequency (see Fig.\,\ref{fig:tsm}b), and for LSMs is a variation of the vacuum-Rabi splitting $2g_{cm}$.
Since the working band of the two magnetometers is controlled by the dynamics of CMPs, the latter was studied in a separate work\cite{crescini2020magnon}.
%}

In conclusion, we described and operated two different types of CMP-based magnetometers, which show an outstanding magnetic sensitivity.
The TSM is a device that benefits from its scalability, which lowers the minimum detectable magnetization oscillations. We believe that it is more suitable for studying fundamental physics, for instance in the search for Axions, where Dark Matter can be described as a wide, uniform, and persistent rf field acting on the electron spins.
The LSM is a device of simpler application, as it can precisely detect faint magnetic fields localised on a small spin ensemble. Its sensitivity relies on the design and engineering of the PMHS, which can be further developed to reach remarkable sensitivity improvements. We mention that its usage to search for Axions is immediate, and that a single LSM can scan a broad Axion-mass range by changing the CMP vacuum-Rabi splitting.
We showed that the unique features of PMHS makes them suitable to assess fundamental physics problems, and we envision more future applications of these systems as testbeds for precision magnetometry.

\begin{acknowledgments}
The authors would like to acknowledge the contribution of the QUAX collaboration in the development of these devices. We also thank Enrico Berto, Andrea Benato, Fulvio Calaon, and Mario Tessaro for the help in the building of the experimental setups, and in particular for the aid with the mechanics, cryogenics, and electronics of the apparatuses. We acknowledge the support of INFN-Laboratori Nazionali di Legnaro, for hosting all the experimental setups described in this work, and for the availability of large quantities of liquid helium.
\end{acknowledgments}

\section*{Data availability}
The data supporting the findings of this work are available from the corresponding author upon reasonable request.

\bibliography{hybridMagnetometers}

\begin{thebibliography}{72}%
\makeatletter
\providecommand \@ifxundefined [1]{%
 \@ifx{#1\undefined}
}%
\providecommand \@ifnum [1]{%
 \ifnum #1\expandafter \@firstoftwo
 \else \expandafter \@secondoftwo
 \fi
}%
\providecommand \@ifx [1]{%
 \ifx #1\expandafter \@firstoftwo
 \else \expandafter \@secondoftwo
 \fi
}%
\providecommand \natexlab [1]{#1}%
\providecommand \enquote  [1]{``#1''}%
\providecommand \bibnamefont  [1]{#1}%
\providecommand \bibfnamefont [1]{#1}%
\providecommand \citenamefont [1]{#1}%
\providecommand \href@noop [0]{\@secondoftwo}%
\providecommand \href [0]{\begingroup \@sanitize@url \@href}%
\providecommand \@href[1]{\@@startlink{#1}\@@href}%
\providecommand \@@href[1]{\endgroup#1\@@endlink}%
\providecommand \@sanitize@url [0]{\catcode `\\12\catcode `\$12\catcode
  `\&12\catcode `\#12\catcode `\^12\catcode `\_12\catcode `\%12\relax}%
\providecommand \@@startlink[1]{}%
\providecommand \@@endlink[0]{}%
\providecommand \url  [0]{\begingroup\@sanitize@url \@url }%
\providecommand \@url [1]{\endgroup\@href {#1}{\urlprefix }}%
\providecommand \urlprefix  [0]{URL }%
\providecommand \Eprint [0]{\href }%
\providecommand \doibase [0]{http://dx.doi.org/}%
\providecommand \selectlanguage [0]{\@gobble}%
\providecommand \bibinfo  [0]{\@secondoftwo}%
\providecommand \bibfield  [0]{\@secondoftwo}%
\providecommand \translation [1]{[#1]}%
\providecommand \BibitemOpen [0]{}%
\providecommand \bibitemStop [0]{}%
\providecommand \bibitemNoStop [0]{.\EOS\space}%
\providecommand \EOS [0]{\spacefactor3000\relax}%
\providecommand \BibitemShut  [1]{\csname bibitem#1\endcsname}%
\let\auto@bib@innerbib\@empty
%</preamble>
\bibitem [{\citenamefont {Clerk}\ \emph {et~al.}(2020)\citenamefont {Clerk},
  \citenamefont {Lehnert}, \citenamefont {Bertet}, \citenamefont {Petta},\ and\
  \citenamefont {Nakamura}}]{Clerk2020}%
  \BibitemOpen
  \bibfield  {author} {\bibinfo {author} {\bibfnamefont {A.~A.}\ \bibnamefont
  {Clerk}}, \bibinfo {author} {\bibfnamefont {K.~W.}\ \bibnamefont {Lehnert}},
  \bibinfo {author} {\bibfnamefont {P.}~\bibnamefont {Bertet}}, \bibinfo
  {author} {\bibfnamefont {J.~R.}\ \bibnamefont {Petta}}, \ and\ \bibinfo
  {author} {\bibfnamefont {Y.}~\bibnamefont {Nakamura}},\ }\bibfield  {title}
  {\enquote {\bibinfo {title} {Hybrid quantum systems with circuit quantum
  electrodynamics},}\ }\href {\doibase 10.1038/s41567-020-0797-9} {\bibfield
  {journal} {\bibinfo  {journal} {Nature Physics}\ }\textbf {\bibinfo {volume}
  {16}},\ \bibinfo {pages} {257--267} (\bibinfo {year} {2020})}\BibitemShut
  {NoStop}%
\bibitem [{\citenamefont {Lachance-Quirion}\ \emph {et~al.}(2019)\citenamefont
  {Lachance-Quirion}, \citenamefont {Tabuchi}, \citenamefont {Gloppe},
  \citenamefont {Usami},\ and\ \citenamefont
  {Nakamura}}]{Lachance_Quirion_2019}%
  \BibitemOpen
  \bibfield  {author} {\bibinfo {author} {\bibfnamefont {D.}~\bibnamefont
  {Lachance-Quirion}}, \bibinfo {author} {\bibfnamefont {Y.}~\bibnamefont
  {Tabuchi}}, \bibinfo {author} {\bibfnamefont {A.}~\bibnamefont {Gloppe}},
  \bibinfo {author} {\bibfnamefont {K.}~\bibnamefont {Usami}}, \ and\ \bibinfo
  {author} {\bibfnamefont {Y.}~\bibnamefont {Nakamura}},\ }\bibfield  {title}
  {\enquote {\bibinfo {title} {Hybrid quantum systems based on magnonics},}\
  }\href {\doibase 10.7567/1882-0786/ab248d} {\bibfield  {journal} {\bibinfo
  {journal} {Applied Physics Express}\ }\textbf {\bibinfo {volume} {12}},\
  \bibinfo {pages} {070101} (\bibinfo {year} {2019})}\BibitemShut {NoStop}%
\bibitem [{\citenamefont {Haroche}\ and\ \citenamefont
  {Raimond}(2006)}]{haroche}%
  \BibitemOpen
  \bibfield  {author} {\bibinfo {author} {\bibfnamefont {S.}~\bibnamefont
  {Haroche}}\ and\ \bibinfo {author} {\bibfnamefont {J.-M.}\ \bibnamefont
  {Raimond}},\ }\href@noop {} {\emph {\bibinfo {title} {Exploring the Quantum:
  Atoms, Cavities, and Photons}}}\ (\bibinfo  {publisher} {Oxford University
  Press},\ \bibinfo {year} {2006})\BibitemShut {NoStop}%
\bibitem [{\citenamefont {Chumak}\ \emph {et~al.}(2015)\citenamefont {Chumak},
  \citenamefont {Vasyuchka}, \citenamefont {Serga},\ and\ \citenamefont
  {Hillebrands}}]{Chumak2015}%
  \BibitemOpen
  \bibfield  {author} {\bibinfo {author} {\bibfnamefont {A.~V.}\ \bibnamefont
  {Chumak}}, \bibinfo {author} {\bibfnamefont {V.~I.}\ \bibnamefont
  {Vasyuchka}}, \bibinfo {author} {\bibfnamefont {A.~A.}\ \bibnamefont
  {Serga}}, \ and\ \bibinfo {author} {\bibfnamefont {B.}~\bibnamefont
  {Hillebrands}},\ }\bibfield  {title} {\enquote {\bibinfo {title} {Magnon
  spintronics},}\ }\href {\doibase 10.1038/nphys3347} {\bibfield  {journal}
  {\bibinfo  {journal} {Nature Physics}\ }\textbf {\bibinfo {volume} {11}},\
  \bibinfo {pages} {453--461} (\bibinfo {year} {2015})}\BibitemShut {NoStop}%
\bibitem [{\citenamefont {Degen}, \citenamefont {Reinhard},\ and\ \citenamefont
  {Cappellaro}(2017)}]{RevModPhys.89.035002}%
  \BibitemOpen
  \bibfield  {author} {\bibinfo {author} {\bibfnamefont {C.~L.}\ \bibnamefont
  {Degen}}, \bibinfo {author} {\bibfnamefont {F.}~\bibnamefont {Reinhard}}, \
  and\ \bibinfo {author} {\bibfnamefont {P.}~\bibnamefont {Cappellaro}},\
  }\bibfield  {title} {\enquote {\bibinfo {title} {Quantum sensing},}\ }\href
  {\doibase 10.1103/RevModPhys.89.035002} {\bibfield  {journal} {\bibinfo
  {journal} {Rev. Mod. Phys.}\ }\textbf {\bibinfo {volume} {89}},\ \bibinfo
  {pages} {035002} (\bibinfo {year} {2017})}\BibitemShut {NoStop}%
\bibitem [{\citenamefont {Kurizki}\ \emph {et~al.}(2015)\citenamefont
  {Kurizki}, \citenamefont {Bertet}, \citenamefont {Kubo}, \citenamefont
  {M{\o}lmer}, \citenamefont {Petrosyan}, \citenamefont {Rabl},\ and\
  \citenamefont {Schmiedmayer}}]{Kurizki3866}%
  \BibitemOpen
  \bibfield  {author} {\bibinfo {author} {\bibfnamefont {G.}~\bibnamefont
  {Kurizki}}, \bibinfo {author} {\bibfnamefont {P.}~\bibnamefont {Bertet}},
  \bibinfo {author} {\bibfnamefont {Y.}~\bibnamefont {Kubo}}, \bibinfo {author}
  {\bibfnamefont {K.}~\bibnamefont {M{\o}lmer}}, \bibinfo {author}
  {\bibfnamefont {D.}~\bibnamefont {Petrosyan}}, \bibinfo {author}
  {\bibfnamefont {P.}~\bibnamefont {Rabl}}, \ and\ \bibinfo {author}
  {\bibfnamefont {J.}~\bibnamefont {Schmiedmayer}},\ }\bibfield  {title}
  {\enquote {\bibinfo {title} {Quantum technologies with hybrid systems},}\
  }\href {\doibase 10.1073/pnas.1419326112} {\bibfield  {journal} {\bibinfo
  {journal} {Proceedings of the National Academy of Sciences}\ }\textbf
  {\bibinfo {volume} {112}},\ \bibinfo {pages} {3866--3873} (\bibinfo {year}
  {2015})},\ \Eprint
  {http://arxiv.org/abs/http://www.pnas.org/content/112/13/3866.full.pdf}
  {http://www.pnas.org/content/112/13/3866.full.pdf} \BibitemShut {NoStop}%
\bibitem [{\citenamefont {Shen}\ and\ \citenamefont
  {Bloembergen}(1966)}]{PhysRev.143.372}%
  \BibitemOpen
  \bibfield  {author} {\bibinfo {author} {\bibfnamefont {Y.~R.}\ \bibnamefont
  {Shen}}\ and\ \bibinfo {author} {\bibfnamefont {N.}~\bibnamefont
  {Bloembergen}},\ }\bibfield  {title} {\enquote {\bibinfo {title} {Interaction
  between light waves and spin waves},}\ }\href {\doibase
  10.1103/PhysRev.143.372} {\bibfield  {journal} {\bibinfo  {journal} {Phys.
  Rev.}\ }\textbf {\bibinfo {volume} {143}},\ \bibinfo {pages} {372--384}
  (\bibinfo {year} {1966})}\BibitemShut {NoStop}%
\bibitem [{\citenamefont {Kittel}(2004)}]{Kittel2004}%
  \BibitemOpen
  \bibfield  {author} {\bibinfo {author} {\bibfnamefont {C.}~\bibnamefont
  {Kittel}},\ }\href
  {http://www.amazon.com/Introduction-Solid-Physics-Charles-Kittel/dp/047141526X/ref=dp_ob_title_bk}
  {\emph {\bibinfo {title} {Introduction to Solid State Physics}}},\ \bibinfo
  {edition} {8th}\ ed.\ (\bibinfo  {publisher} {Wiley},\ \bibinfo {year}
  {2004})\BibitemShut {NoStop}%
\bibitem [{\citenamefont {Walls}\ and\ \citenamefont
  {Milburn}(2007)}]{walls2007quantum}%
  \BibitemOpen
  \bibfield  {author} {\bibinfo {author} {\bibfnamefont {D.~F.}\ \bibnamefont
  {Walls}}\ and\ \bibinfo {author} {\bibfnamefont {G.~J.}\ \bibnamefont
  {Milburn}},\ }\href@noop {} {\emph {\bibinfo {title} {Quantum optics}}}\
  (\bibinfo  {publisher} {Springer Science \& Business Media},\ \bibinfo {year}
  {2007})\BibitemShut {NoStop}%
\bibitem [{\citenamefont {Tabuchi}\ \emph {et~al.}(2016)\citenamefont
  {Tabuchi}, \citenamefont {Ishino}, \citenamefont {Noguchi}, \citenamefont
  {Ishikawa}, \citenamefont {Yamazaki}, \citenamefont {Usami},\ and\
  \citenamefont {Nakamura}}]{TABUCHI2016729}%
  \BibitemOpen
  \bibfield  {author} {\bibinfo {author} {\bibfnamefont {Y.}~\bibnamefont
  {Tabuchi}}, \bibinfo {author} {\bibfnamefont {S.}~\bibnamefont {Ishino}},
  \bibinfo {author} {\bibfnamefont {A.}~\bibnamefont {Noguchi}}, \bibinfo
  {author} {\bibfnamefont {T.}~\bibnamefont {Ishikawa}}, \bibinfo {author}
  {\bibfnamefont {R.}~\bibnamefont {Yamazaki}}, \bibinfo {author}
  {\bibfnamefont {K.}~\bibnamefont {Usami}}, \ and\ \bibinfo {author}
  {\bibfnamefont {Y.}~\bibnamefont {Nakamura}},\ }\bibfield  {title} {\enquote
  {\bibinfo {title} {Quantum magnonics: The magnon meets the superconducting
  qubit},}\ }\href {\doibase https://doi.org/10.1016/j.crhy.2016.07.009}
  {\bibfield  {journal} {\bibinfo  {journal} {Comptes Rendus Physique}\
  }\textbf {\bibinfo {volume} {17}},\ \bibinfo {pages} {729 -- 739} (\bibinfo
  {year} {2016})},\ \bibinfo {note} {quantum microwaves / Micro-ondes
  quantiques}\BibitemShut {NoStop}%
\bibitem [{\citenamefont {Zhang}\ \emph
  {et~al.}(2015{\natexlab{a}})\citenamefont {Zhang}, \citenamefont {Zou},
  \citenamefont {Zhu}, \citenamefont {Marquardt}, \citenamefont {Jiang},\ and\
  \citenamefont {Tang}}]{Zhang2015b}%
  \BibitemOpen
  \bibfield  {author} {\bibinfo {author} {\bibfnamefont {X.}~\bibnamefont
  {Zhang}}, \bibinfo {author} {\bibfnamefont {C.-L.}\ \bibnamefont {Zou}},
  \bibinfo {author} {\bibfnamefont {N.}~\bibnamefont {Zhu}}, \bibinfo {author}
  {\bibfnamefont {F.}~\bibnamefont {Marquardt}}, \bibinfo {author}
  {\bibfnamefont {L.}~\bibnamefont {Jiang}}, \ and\ \bibinfo {author}
  {\bibfnamefont {H.~X.}\ \bibnamefont {Tang}},\ }\bibfield  {title} {\enquote
  {\bibinfo {title} {Magnon dark modes and gradient memory},}\ }\href {\doibase
  10.1038/ncomms9914} {\bibfield  {journal} {\bibinfo  {journal} {Nature
  Communications}\ }\textbf {\bibinfo {volume} {6}},\ \bibinfo {pages} {8914}
  (\bibinfo {year} {2015}{\natexlab{a}})}\BibitemShut {NoStop}%
\bibitem [{\citenamefont {Schuster}\ \emph {et~al.}(2010)\citenamefont
  {Schuster}, \citenamefont {Sears}, \citenamefont {Ginossar}, \citenamefont
  {DiCarlo}, \citenamefont {Frunzio}, \citenamefont {Morton}, \citenamefont
  {Wu}, \citenamefont {Briggs}, \citenamefont {Buckley}, \citenamefont
  {Awschalom},\ and\ \citenamefont {Schoelkopf}}]{PhysRevLett.105.140501}%
  \BibitemOpen
  \bibfield  {author} {\bibinfo {author} {\bibfnamefont {D.~I.}\ \bibnamefont
  {Schuster}}, \bibinfo {author} {\bibfnamefont {A.~P.}\ \bibnamefont {Sears}},
  \bibinfo {author} {\bibfnamefont {E.}~\bibnamefont {Ginossar}}, \bibinfo
  {author} {\bibfnamefont {L.}~\bibnamefont {DiCarlo}}, \bibinfo {author}
  {\bibfnamefont {L.}~\bibnamefont {Frunzio}}, \bibinfo {author} {\bibfnamefont
  {J.~J.~L.}\ \bibnamefont {Morton}}, \bibinfo {author} {\bibfnamefont
  {H.}~\bibnamefont {Wu}}, \bibinfo {author} {\bibfnamefont {G.~A.~D.}\
  \bibnamefont {Briggs}}, \bibinfo {author} {\bibfnamefont {B.~B.}\
  \bibnamefont {Buckley}}, \bibinfo {author} {\bibfnamefont {D.~D.}\
  \bibnamefont {Awschalom}}, \ and\ \bibinfo {author} {\bibfnamefont {R.~J.}\
  \bibnamefont {Schoelkopf}},\ }\bibfield  {title} {\enquote {\bibinfo {title}
  {High-cooperativity coupling of electron-spin ensembles to superconducting
  cavities},}\ }\href {\doibase 10.1103/PhysRevLett.105.140501} {\bibfield
  {journal} {\bibinfo  {journal} {Phys. Rev. Lett.}\ }\textbf {\bibinfo
  {volume} {105}},\ \bibinfo {pages} {140501} (\bibinfo {year}
  {2010})}\BibitemShut {NoStop}%
\bibitem [{\citenamefont {Ghirri}\ \emph {et~al.}(2015)\citenamefont {Ghirri},
  \citenamefont {Bonizzoni}, \citenamefont {Gerace}, \citenamefont {Sanna},
  \citenamefont {Cassinese},\ and\ \citenamefont {Affronte}}]{bonizzoni}%
  \BibitemOpen
  \bibfield  {author} {\bibinfo {author} {\bibfnamefont {A.}~\bibnamefont
  {Ghirri}}, \bibinfo {author} {\bibfnamefont {C.}~\bibnamefont {Bonizzoni}},
  \bibinfo {author} {\bibfnamefont {D.}~\bibnamefont {Gerace}}, \bibinfo
  {author} {\bibfnamefont {S.}~\bibnamefont {Sanna}}, \bibinfo {author}
  {\bibfnamefont {A.}~\bibnamefont {Cassinese}}, \ and\ \bibinfo {author}
  {\bibfnamefont {M.}~\bibnamefont {Affronte}},\ }\bibfield  {title} {\enquote
  {\bibinfo {title} {Ybco microwave resonators for strong collective coupling
  with spin ensembles},}\ }\bibfield  {booktitle} {\emph {\bibinfo {booktitle}
  {Applied Physics Letters}},\ }\href@noop {} {\ \textbf {\bibinfo {volume}
  {106}} (\bibinfo {year} {2015})}\BibitemShut {NoStop}%
\bibitem [{\citenamefont {Soykal}\ and\ \citenamefont
  {Flatt\'e}(2010)}]{PhysRevLett.104.077202}%
  \BibitemOpen
  \bibfield  {author} {\bibinfo {author} {\bibfnamefont {O.~O.}\ \bibnamefont
  {Soykal}}\ and\ \bibinfo {author} {\bibfnamefont {M.~E.}\ \bibnamefont
  {Flatt\'e}},\ }\bibfield  {title} {\enquote {\bibinfo {title} {Strong field
  interactions between a nanomagnet and a photonic cavity},}\ }\href {\doibase
  10.1103/PhysRevLett.104.077202} {\bibfield  {journal} {\bibinfo  {journal}
  {Phys. Rev. Lett.}\ }\textbf {\bibinfo {volume} {104}},\ \bibinfo {pages}
  {077202} (\bibinfo {year} {2010})}\BibitemShut {NoStop}%
\bibitem [{\citenamefont {Ghirri}\ \emph {et~al.}(2016)\citenamefont {Ghirri},
  \citenamefont {Bonizzoni}, \citenamefont {Troiani}, \citenamefont {Buccheri},
  \citenamefont {Beverina}, \citenamefont {Cassinese},\ and\ \citenamefont
  {Affronte}}]{bonizzoni2}%
  \BibitemOpen
  \bibfield  {author} {\bibinfo {author} {\bibfnamefont {A.}~\bibnamefont
  {Ghirri}}, \bibinfo {author} {\bibfnamefont {C.}~\bibnamefont {Bonizzoni}},
  \bibinfo {author} {\bibfnamefont {F.}~\bibnamefont {Troiani}}, \bibinfo
  {author} {\bibfnamefont {N.}~\bibnamefont {Buccheri}}, \bibinfo {author}
  {\bibfnamefont {L.}~\bibnamefont {Beverina}}, \bibinfo {author}
  {\bibfnamefont {A.}~\bibnamefont {Cassinese}}, \ and\ \bibinfo {author}
  {\bibfnamefont {M.}~\bibnamefont {Affronte}},\ }\bibfield  {title} {\enquote
  {\bibinfo {title} {Coherently coupling distinct spin ensembles through a
  high-$t_c$ superconducting resonator},}\ }\bibfield  {booktitle} {\emph
  {\bibinfo {booktitle} {Physical Review A}},\ }\href@noop {} {\ \textbf
  {\bibinfo {volume} {93}} (\bibinfo {year} {2016})}\BibitemShut {NoStop}%
\bibitem [{\citenamefont {Heshami}\ \emph {et~al.}(2016)\citenamefont
  {Heshami}, \citenamefont {England}, \citenamefont {Humphreys}, \citenamefont
  {Bustard}, \citenamefont {Acosta}, \citenamefont {Nunn},\ and\ \citenamefont
  {Sussman}}]{doi:10.1080/09500340.2016.1148212}%
  \BibitemOpen
  \bibfield  {author} {\bibinfo {author} {\bibfnamefont {K.}~\bibnamefont
  {Heshami}}, \bibinfo {author} {\bibfnamefont {D.~G.}\ \bibnamefont
  {England}}, \bibinfo {author} {\bibfnamefont {P.~C.}\ \bibnamefont
  {Humphreys}}, \bibinfo {author} {\bibfnamefont {P.~J.}\ \bibnamefont
  {Bustard}}, \bibinfo {author} {\bibfnamefont {V.~M.}\ \bibnamefont {Acosta}},
  \bibinfo {author} {\bibfnamefont {J.}~\bibnamefont {Nunn}}, \ and\ \bibinfo
  {author} {\bibfnamefont {B.~J.}\ \bibnamefont {Sussman}},\ }\bibfield
  {title} {\enquote {\bibinfo {title} {Quantum memories: emerging applications
  and recent advances},}\ }\href {\doibase 10.1080/09500340.2016.1148212}
  {\bibfield  {journal} {\bibinfo  {journal} {Journal of Modern Optics}\
  }\textbf {\bibinfo {volume} {63}},\ \bibinfo {pages} {2005--2028} (\bibinfo
  {year} {2016})},\ \Eprint
  {http://arxiv.org/abs/https://doi.org/10.1080/09500340.2016.1148212}
  {https://doi.org/10.1080/09500340.2016.1148212} \BibitemShut {NoStop}%
\bibitem [{\citenamefont {Kimble}(2008)}]{Kimble2008}%
  \BibitemOpen
  \bibfield  {author} {\bibinfo {author} {\bibfnamefont {H.~J.}\ \bibnamefont
  {Kimble}},\ }\bibfield  {title} {\enquote {\bibinfo {title} {The quantum
  internet},}\ }\href {\doibase 10.1038/nature07127} {\bibfield  {journal}
  {\bibinfo  {journal} {Nature}\ }\textbf {\bibinfo {volume} {453}},\ \bibinfo
  {pages} {1023--1030} (\bibinfo {year} {2008})}\BibitemShut {NoStop}%
\bibitem [{\citenamefont {Williamson}, \citenamefont {Chen},\ and\
  \citenamefont {Longdell}(2014)}]{PhysRevLett.113.203601}%
  \BibitemOpen
  \bibfield  {author} {\bibinfo {author} {\bibfnamefont {L.~A.}\ \bibnamefont
  {Williamson}}, \bibinfo {author} {\bibfnamefont {Y.-H.}\ \bibnamefont
  {Chen}}, \ and\ \bibinfo {author} {\bibfnamefont {J.~J.}\ \bibnamefont
  {Longdell}},\ }\bibfield  {title} {\enquote {\bibinfo {title} {Magneto-optic
  modulator with unit quantum efficiency},}\ }\href {\doibase
  10.1103/PhysRevLett.113.203601} {\bibfield  {journal} {\bibinfo  {journal}
  {Phys. Rev. Lett.}\ }\textbf {\bibinfo {volume} {113}},\ \bibinfo {pages}
  {203601} (\bibinfo {year} {2014})}\BibitemShut {NoStop}%
\bibitem [{\citenamefont {Fernandez-Gonzalvo}\ \emph
  {et~al.}(2015)\citenamefont {Fernandez-Gonzalvo}, \citenamefont {Chen},
  \citenamefont {Yin}, \citenamefont {Rogge},\ and\ \citenamefont
  {Longdell}}]{PhysRevA.92.062313}%
  \BibitemOpen
  \bibfield  {author} {\bibinfo {author} {\bibfnamefont {X.}~\bibnamefont
  {Fernandez-Gonzalvo}}, \bibinfo {author} {\bibfnamefont {Y.-H.}\ \bibnamefont
  {Chen}}, \bibinfo {author} {\bibfnamefont {C.}~\bibnamefont {Yin}}, \bibinfo
  {author} {\bibfnamefont {S.}~\bibnamefont {Rogge}}, \ and\ \bibinfo {author}
  {\bibfnamefont {J.~J.}\ \bibnamefont {Longdell}},\ }\bibfield  {title}
  {\enquote {\bibinfo {title} {Coherent frequency up-conversion of microwaves
  to the optical telecommunications band in an er:yso crystal},}\ }\href
  {\doibase 10.1103/PhysRevA.92.062313} {\bibfield  {journal} {\bibinfo
  {journal} {Phys. Rev. A}\ }\textbf {\bibinfo {volume} {92}},\ \bibinfo
  {pages} {062313} (\bibinfo {year} {2015})}\BibitemShut {NoStop}%
\bibitem [{\citenamefont {Hisatomi}\ \emph {et~al.}(2016)\citenamefont
  {Hisatomi}, \citenamefont {Osada}, \citenamefont {Tabuchi}, \citenamefont
  {Ishikawa}, \citenamefont {Noguchi}, \citenamefont {Yamazaki}, \citenamefont
  {Usami},\ and\ \citenamefont {Nakamura}}]{PhysRevB.93.174427}%
  \BibitemOpen
  \bibfield  {author} {\bibinfo {author} {\bibfnamefont {R.}~\bibnamefont
  {Hisatomi}}, \bibinfo {author} {\bibfnamefont {A.}~\bibnamefont {Osada}},
  \bibinfo {author} {\bibfnamefont {Y.}~\bibnamefont {Tabuchi}}, \bibinfo
  {author} {\bibfnamefont {T.}~\bibnamefont {Ishikawa}}, \bibinfo {author}
  {\bibfnamefont {A.}~\bibnamefont {Noguchi}}, \bibinfo {author} {\bibfnamefont
  {R.}~\bibnamefont {Yamazaki}}, \bibinfo {author} {\bibfnamefont
  {K.}~\bibnamefont {Usami}}, \ and\ \bibinfo {author} {\bibfnamefont
  {Y.}~\bibnamefont {Nakamura}},\ }\bibfield  {title} {\enquote {\bibinfo
  {title} {Bidirectional conversion between microwave and light via
  ferromagnetic magnons},}\ }\href {\doibase 10.1103/PhysRevB.93.174427}
  {\bibfield  {journal} {\bibinfo  {journal} {Phys. Rev. B}\ }\textbf {\bibinfo
  {volume} {93}},\ \bibinfo {pages} {174427} (\bibinfo {year}
  {2016})}\BibitemShut {NoStop}%
\bibitem [{\citenamefont {Braggio}\ \emph {et~al.}(2016)\citenamefont
  {Braggio}, \citenamefont {Carugno}, \citenamefont {Guarise}, \citenamefont
  {Ortolan},\ and\ \citenamefont {Ruoso}}]{cate}%
  \BibitemOpen
  \bibfield  {author} {\bibinfo {author} {\bibfnamefont {C.}~\bibnamefont
  {Braggio}}, \bibinfo {author} {\bibfnamefont {G.}~\bibnamefont {Carugno}},
  \bibinfo {author} {\bibfnamefont {M.}~\bibnamefont {Guarise}}, \bibinfo
  {author} {\bibfnamefont {A.}~\bibnamefont {Ortolan}}, \ and\ \bibinfo
  {author} {\bibfnamefont {G.}~\bibnamefont {Ruoso}},\ }\bibfield  {title}
  {\enquote {\bibinfo {title} {Optical manipulation of a magnon-photon hybrid
  system},}\ }\bibfield  {booktitle} {\emph {\bibinfo {booktitle} {Physical
  Review Letters}},\ }\href@noop {} {\ \textbf {\bibinfo {volume} {118}}
  (\bibinfo {year} {2016})}\BibitemShut {NoStop}%
\bibitem [{\citenamefont {Lachance-Quirion}\ \emph {et~al.}(2017)\citenamefont
  {Lachance-Quirion}, \citenamefont {Tabuchi}, \citenamefont {Ishino},
  \citenamefont {Noguchi}, \citenamefont {Ishikawa}, \citenamefont {Yamazaki},\
  and\ \citenamefont {Nakamura}}]{Lachance-Quirione1603150}%
  \BibitemOpen
  \bibfield  {author} {\bibinfo {author} {\bibfnamefont {D.}~\bibnamefont
  {Lachance-Quirion}}, \bibinfo {author} {\bibfnamefont {Y.}~\bibnamefont
  {Tabuchi}}, \bibinfo {author} {\bibfnamefont {S.}~\bibnamefont {Ishino}},
  \bibinfo {author} {\bibfnamefont {A.}~\bibnamefont {Noguchi}}, \bibinfo
  {author} {\bibfnamefont {T.}~\bibnamefont {Ishikawa}}, \bibinfo {author}
  {\bibfnamefont {R.}~\bibnamefont {Yamazaki}}, \ and\ \bibinfo {author}
  {\bibfnamefont {Y.}~\bibnamefont {Nakamura}},\ }\bibfield  {title} {\enquote
  {\bibinfo {title} {Resolving quanta of collective spin excitations in a
  millimeter-sized ferromagnet},}\ }\href {\doibase 10.1126/sciadv.1603150}
  {\bibfield  {journal} {\bibinfo  {journal} {Science Advances}\ }\textbf
  {\bibinfo {volume} {3}} (\bibinfo {year} {2017}),\ 10.1126/sciadv.1603150},\
  \Eprint
  {http://arxiv.org/abs/https://advances.sciencemag.org/content/3/7/e1603150.full.pdf}
  {https://advances.sciencemag.org/content/3/7/e1603150.full.pdf} \BibitemShut
  {NoStop}%
\bibitem [{\citenamefont {Lachance-Quirion}\ \emph {et~al.}(2020)\citenamefont
  {Lachance-Quirion}, \citenamefont {Wolski}, \citenamefont {Tabuchi},
  \citenamefont {Kono}, \citenamefont {Usami},\ and\ \citenamefont
  {Nakamura}}]{Lachance-Quirion425}%
  \BibitemOpen
  \bibfield  {author} {\bibinfo {author} {\bibfnamefont {D.}~\bibnamefont
  {Lachance-Quirion}}, \bibinfo {author} {\bibfnamefont {S.~P.}\ \bibnamefont
  {Wolski}}, \bibinfo {author} {\bibfnamefont {Y.}~\bibnamefont {Tabuchi}},
  \bibinfo {author} {\bibfnamefont {S.}~\bibnamefont {Kono}}, \bibinfo {author}
  {\bibfnamefont {K.}~\bibnamefont {Usami}}, \ and\ \bibinfo {author}
  {\bibfnamefont {Y.}~\bibnamefont {Nakamura}},\ }\bibfield  {title} {\enquote
  {\bibinfo {title} {Entanglement-based single-shot detection of a single
  magnon with a superconducting qubit},}\ }\href {\doibase
  10.1126/science.aaz9236} {\bibfield  {journal} {\bibinfo  {journal}
  {Science}\ }\textbf {\bibinfo {volume} {367}},\ \bibinfo {pages} {425--428}
  (\bibinfo {year} {2020})},\ \Eprint
  {http://arxiv.org/abs/https://science.sciencemag.org/content/367/6476/425.full.pdf}
  {https://science.sciencemag.org/content/367/6476/425.full.pdf} \BibitemShut
  {NoStop}%
\bibitem [{\citenamefont {Wolski}\ \emph {et~al.}(2020)\citenamefont {Wolski},
  \citenamefont {Lachance-Quirion}, \citenamefont {Tabuchi}, \citenamefont
  {Kono}, \citenamefont {Noguchi}, \citenamefont {Usami},\ and\ \citenamefont
  {Nakamura}}]{wolski2020dissipationbased}%
  \BibitemOpen
  \bibfield  {author} {\bibinfo {author} {\bibfnamefont {S.~P.}\ \bibnamefont
  {Wolski}}, \bibinfo {author} {\bibfnamefont {D.}~\bibnamefont
  {Lachance-Quirion}}, \bibinfo {author} {\bibfnamefont {Y.}~\bibnamefont
  {Tabuchi}}, \bibinfo {author} {\bibfnamefont {S.}~\bibnamefont {Kono}},
  \bibinfo {author} {\bibfnamefont {A.}~\bibnamefont {Noguchi}}, \bibinfo
  {author} {\bibfnamefont {K.}~\bibnamefont {Usami}}, \ and\ \bibinfo {author}
  {\bibfnamefont {Y.}~\bibnamefont {Nakamura}},\ }\href@noop {} {\enquote
  {\bibinfo {title} {Dissipation-based quantum sensing of magnons with a
  superconducting qubit},}\ } (\bibinfo {year} {2020}),\ \Eprint
  {http://arxiv.org/abs/2005.09250} {arXiv:2005.09250 [quant-ph]} \BibitemShut
  {NoStop}%
\bibitem [{\citenamefont {Bender}(2007)}]{Bender_2007}%
  \BibitemOpen
  \bibfield  {author} {\bibinfo {author} {\bibfnamefont {C.~M.}\ \bibnamefont
  {Bender}},\ }\bibfield  {title} {\enquote {\bibinfo {title} {Making sense of
  non-hermitian hamiltonians},}\ }\href {\doibase 10.1088/0034-4885/70/6/r03}
  {\bibfield  {journal} {\bibinfo  {journal} {Reports on Progress in Physics}\
  }\textbf {\bibinfo {volume} {70}},\ \bibinfo {pages} {947--1018} (\bibinfo
  {year} {2007})}\BibitemShut {NoStop}%
\bibitem [{\citenamefont {Bender}\ and\ \citenamefont
  {Boettcher}(1998)}]{PhysRevLett.80.5243}%
  \BibitemOpen
  \bibfield  {author} {\bibinfo {author} {\bibfnamefont {C.~M.}\ \bibnamefont
  {Bender}}\ and\ \bibinfo {author} {\bibfnamefont {S.}~\bibnamefont
  {Boettcher}},\ }\bibfield  {title} {\enquote {\bibinfo {title} {Real spectra
  in non-hermitian hamiltonians having $pt$ symmetry},}\ }\href {\doibase
  10.1103/PhysRevLett.80.5243} {\bibfield  {journal} {\bibinfo  {journal}
  {Phys. Rev. Lett.}\ }\textbf {\bibinfo {volume} {80}},\ \bibinfo {pages}
  {5243--5246} (\bibinfo {year} {1998})}\BibitemShut {NoStop}%
\bibitem [{\citenamefont {R{\"u}ter}\ \emph {et~al.}(2010)\citenamefont
  {R{\"u}ter}, \citenamefont {Makris}, \citenamefont {El-Ganainy},
  \citenamefont {Christodoulides}, \citenamefont {Segev},\ and\ \citenamefont
  {Kip}}]{Ruter2010}%
  \BibitemOpen
  \bibfield  {author} {\bibinfo {author} {\bibfnamefont {C.~E.}\ \bibnamefont
  {R{\"u}ter}}, \bibinfo {author} {\bibfnamefont {K.~G.}\ \bibnamefont
  {Makris}}, \bibinfo {author} {\bibfnamefont {R.}~\bibnamefont {El-Ganainy}},
  \bibinfo {author} {\bibfnamefont {D.~N.}\ \bibnamefont {Christodoulides}},
  \bibinfo {author} {\bibfnamefont {M.}~\bibnamefont {Segev}}, \ and\ \bibinfo
  {author} {\bibfnamefont {D.}~\bibnamefont {Kip}},\ }\bibfield  {title}
  {\enquote {\bibinfo {title} {Observation of parity--time symmetry in
  optics},}\ }\href {\doibase 10.1038/nphys1515} {\bibfield  {journal}
  {\bibinfo  {journal} {Nature Physics}\ }\textbf {\bibinfo {volume} {6}},\
  \bibinfo {pages} {192--195} (\bibinfo {year} {2010})}\BibitemShut {NoStop}%
\bibitem [{\citenamefont {Zhang}\ \emph {et~al.}(2017)\citenamefont {Zhang},
  \citenamefont {Luo}, \citenamefont {Wang}, \citenamefont {Li},\ and\
  \citenamefont {You}}]{Zhang2017}%
  \BibitemOpen
  \bibfield  {author} {\bibinfo {author} {\bibfnamefont {D.}~\bibnamefont
  {Zhang}}, \bibinfo {author} {\bibfnamefont {X.-Q.}\ \bibnamefont {Luo}},
  \bibinfo {author} {\bibfnamefont {Y.-P.}\ \bibnamefont {Wang}}, \bibinfo
  {author} {\bibfnamefont {T.-F.}\ \bibnamefont {Li}}, \ and\ \bibinfo {author}
  {\bibfnamefont {J.~Q.}\ \bibnamefont {You}},\ }\bibfield  {title} {\enquote
  {\bibinfo {title} {Observation of the exceptional point in cavity
  magnon-polaritons},}\ }\href {\doibase 10.1038/s41467-017-01634-w} {\bibfield
   {journal} {\bibinfo  {journal} {Nature Communications}\ }\textbf {\bibinfo
  {volume} {8}},\ \bibinfo {pages} {1368} (\bibinfo {year} {2017})}\BibitemShut
  {NoStop}%
\bibitem [{\citenamefont {Zhang}\ \emph {et~al.}(2019)\citenamefont {Zhang},
  \citenamefont {Ding}, \citenamefont {Zhou}, \citenamefont {Xu},\ and\
  \citenamefont {Jin}}]{PhysRevLett.123.237202}%
  \BibitemOpen
  \bibfield  {author} {\bibinfo {author} {\bibfnamefont {X.}~\bibnamefont
  {Zhang}}, \bibinfo {author} {\bibfnamefont {K.}~\bibnamefont {Ding}},
  \bibinfo {author} {\bibfnamefont {X.}~\bibnamefont {Zhou}}, \bibinfo {author}
  {\bibfnamefont {J.}~\bibnamefont {Xu}}, \ and\ \bibinfo {author}
  {\bibfnamefont {D.}~\bibnamefont {Jin}},\ }\bibfield  {title} {\enquote
  {\bibinfo {title} {Experimental observation of an exceptional surface in
  synthetic dimensions with magnon polaritons},}\ }\href {\doibase
  10.1103/PhysRevLett.123.237202} {\bibfield  {journal} {\bibinfo  {journal}
  {Phys. Rev. Lett.}\ }\textbf {\bibinfo {volume} {123}},\ \bibinfo {pages}
  {237202} (\bibinfo {year} {2019})}\BibitemShut {NoStop}%
\bibitem [{\citenamefont {Ding}\ \emph {et~al.}(2016)\citenamefont {Ding},
  \citenamefont {Ma}, \citenamefont {Xiao}, \citenamefont {Zhang},\ and\
  \citenamefont {Chan}}]{PhysRevX.6.021007}%
  \BibitemOpen
  \bibfield  {author} {\bibinfo {author} {\bibfnamefont {K.}~\bibnamefont
  {Ding}}, \bibinfo {author} {\bibfnamefont {G.}~\bibnamefont {Ma}}, \bibinfo
  {author} {\bibfnamefont {M.}~\bibnamefont {Xiao}}, \bibinfo {author}
  {\bibfnamefont {Z.~Q.}\ \bibnamefont {Zhang}}, \ and\ \bibinfo {author}
  {\bibfnamefont {C.~T.}\ \bibnamefont {Chan}},\ }\bibfield  {title} {\enquote
  {\bibinfo {title} {Emergence, coalescence, and topological properties of
  multiple exceptional points and their experimental realization},}\ }\href
  {\doibase 10.1103/PhysRevX.6.021007} {\bibfield  {journal} {\bibinfo
  {journal} {Phys. Rev. X}\ }\textbf {\bibinfo {volume} {6}},\ \bibinfo {pages}
  {021007} (\bibinfo {year} {2016})}\BibitemShut {NoStop}%
\bibitem [{\citenamefont {Zhang}\ and\ \citenamefont
  {You}(2019)}]{PhysRevB.99.054404}%
  \BibitemOpen
  \bibfield  {author} {\bibinfo {author} {\bibfnamefont {G.-Q.}\ \bibnamefont
  {Zhang}}\ and\ \bibinfo {author} {\bibfnamefont {J.~Q.}\ \bibnamefont
  {You}},\ }\bibfield  {title} {\enquote {\bibinfo {title} {Higher-order
  exceptional point in a cavity magnonics system},}\ }\href {\doibase
  10.1103/PhysRevB.99.054404} {\bibfield  {journal} {\bibinfo  {journal} {Phys.
  Rev. B}\ }\textbf {\bibinfo {volume} {99}},\ \bibinfo {pages} {054404}
  (\bibinfo {year} {2019})}\BibitemShut {NoStop}%
\bibitem [{\citenamefont {Cao}\ and\ \citenamefont
  {Yan}(2019)}]{PhysRevB.99.214415}%
  \BibitemOpen
  \bibfield  {author} {\bibinfo {author} {\bibfnamefont {Y.}~\bibnamefont
  {Cao}}\ and\ \bibinfo {author} {\bibfnamefont {P.}~\bibnamefont {Yan}},\
  }\bibfield  {title} {\enquote {\bibinfo {title} {Exceptional magnetic
  sensitivity of $\mathcal{P}\mathcal{T}$-symmetric cavity magnon
  polaritons},}\ }\href {\doibase 10.1103/PhysRevB.99.214415} {\bibfield
  {journal} {\bibinfo  {journal} {Phys. Rev. B}\ }\textbf {\bibinfo {volume}
  {99}},\ \bibinfo {pages} {214415} (\bibinfo {year} {2019})}\BibitemShut
  {NoStop}%
\bibitem [{\citenamefont {Rao}\ \emph {et~al.}(2019)\citenamefont {Rao},
  \citenamefont {Kaur}, \citenamefont {Yao}, \citenamefont {Edwards},
  \citenamefont {Zhao}, \citenamefont {Fan}, \citenamefont {Xue}, \citenamefont
  {Silva}, \citenamefont {Gui},\ and\ \citenamefont {Hu}}]{Rao2019}%
  \BibitemOpen
  \bibfield  {author} {\bibinfo {author} {\bibfnamefont {J.~W.}\ \bibnamefont
  {Rao}}, \bibinfo {author} {\bibfnamefont {S.}~\bibnamefont {Kaur}}, \bibinfo
  {author} {\bibfnamefont {B.~M.}\ \bibnamefont {Yao}}, \bibinfo {author}
  {\bibfnamefont {E.~R.~J.}\ \bibnamefont {Edwards}}, \bibinfo {author}
  {\bibfnamefont {Y.~T.}\ \bibnamefont {Zhao}}, \bibinfo {author}
  {\bibfnamefont {X.}~\bibnamefont {Fan}}, \bibinfo {author} {\bibfnamefont
  {D.}~\bibnamefont {Xue}}, \bibinfo {author} {\bibfnamefont {T.~J.}\
  \bibnamefont {Silva}}, \bibinfo {author} {\bibfnamefont {Y.~S.}\ \bibnamefont
  {Gui}}, \ and\ \bibinfo {author} {\bibfnamefont {C.-M.}\ \bibnamefont {Hu}},\
  }\bibfield  {title} {\enquote {\bibinfo {title} {Analogue of dynamic hall
  effect in cavity magnon polariton system and coherently controlled logic
  device},}\ }\href {\doibase 10.1038/s41467-019-11021-2} {\bibfield  {journal}
  {\bibinfo  {journal} {Nature Communications}\ }\textbf {\bibinfo {volume}
  {10}},\ \bibinfo {pages} {2934} (\bibinfo {year} {2019})}\BibitemShut
  {NoStop}%
\bibitem [{\citenamefont {Lambert}\ \emph {et~al.}(2016)\citenamefont
  {Lambert}, \citenamefont {Haigh}, \citenamefont {Langenfeld}, \citenamefont
  {Doherty},\ and\ \citenamefont {Ferguson}}]{PhysRevA.93.021803}%
  \BibitemOpen
  \bibfield  {author} {\bibinfo {author} {\bibfnamefont {N.~J.}\ \bibnamefont
  {Lambert}}, \bibinfo {author} {\bibfnamefont {J.~A.}\ \bibnamefont {Haigh}},
  \bibinfo {author} {\bibfnamefont {S.}~\bibnamefont {Langenfeld}}, \bibinfo
  {author} {\bibfnamefont {A.~C.}\ \bibnamefont {Doherty}}, \ and\ \bibinfo
  {author} {\bibfnamefont {A.~J.}\ \bibnamefont {Ferguson}},\ }\bibfield
  {title} {\enquote {\bibinfo {title} {Cavity-mediated coherent coupling of
  magnetic moments},}\ }\href {\doibase 10.1103/PhysRevA.93.021803} {\bibfield
  {journal} {\bibinfo  {journal} {Phys. Rev. A}\ }\textbf {\bibinfo {volume}
  {93}},\ \bibinfo {pages} {021803} (\bibinfo {year} {2016})}\BibitemShut
  {NoStop}%
\bibitem [{\citenamefont {Wang}\ \emph {et~al.}(2019)\citenamefont {Wang},
  \citenamefont {Rao}, \citenamefont {Yang}, \citenamefont {Xu}, \citenamefont
  {Gui}, \citenamefont {Yao}, \citenamefont {You},\ and\ \citenamefont
  {Hu}}]{PhysRevLett.123.127202}%
  \BibitemOpen
  \bibfield  {author} {\bibinfo {author} {\bibfnamefont {Y.-P.}\ \bibnamefont
  {Wang}}, \bibinfo {author} {\bibfnamefont {J.~W.}\ \bibnamefont {Rao}},
  \bibinfo {author} {\bibfnamefont {Y.}~\bibnamefont {Yang}}, \bibinfo {author}
  {\bibfnamefont {P.-C.}\ \bibnamefont {Xu}}, \bibinfo {author} {\bibfnamefont
  {Y.~S.}\ \bibnamefont {Gui}}, \bibinfo {author} {\bibfnamefont {B.~M.}\
  \bibnamefont {Yao}}, \bibinfo {author} {\bibfnamefont {J.~Q.}\ \bibnamefont
  {You}}, \ and\ \bibinfo {author} {\bibfnamefont {C.-M.}\ \bibnamefont {Hu}},\
  }\bibfield  {title} {\enquote {\bibinfo {title} {Nonreciprocity and
  unidirectional invisibility in cavity magnonics},}\ }\href {\doibase
  10.1103/PhysRevLett.123.127202} {\bibfield  {journal} {\bibinfo  {journal}
  {Phys. Rev. Lett.}\ }\textbf {\bibinfo {volume} {123}},\ \bibinfo {pages}
  {127202} (\bibinfo {year} {2019})}\BibitemShut {NoStop}%
\bibitem [{\citenamefont {Wang}\ \emph {et~al.}(2018)\citenamefont {Wang},
  \citenamefont {Zhang}, \citenamefont {Zhang}, \citenamefont {Li},
  \citenamefont {Hu},\ and\ \citenamefont {You}}]{PhysRevLett.120.057202}%
  \BibitemOpen
  \bibfield  {author} {\bibinfo {author} {\bibfnamefont {Y.-P.}\ \bibnamefont
  {Wang}}, \bibinfo {author} {\bibfnamefont {G.-Q.}\ \bibnamefont {Zhang}},
  \bibinfo {author} {\bibfnamefont {D.}~\bibnamefont {Zhang}}, \bibinfo
  {author} {\bibfnamefont {T.-F.}\ \bibnamefont {Li}}, \bibinfo {author}
  {\bibfnamefont {C.-M.}\ \bibnamefont {Hu}}, \ and\ \bibinfo {author}
  {\bibfnamefont {J.~Q.}\ \bibnamefont {You}},\ }\bibfield  {title} {\enquote
  {\bibinfo {title} {Bistability of cavity magnon polaritons},}\ }\href
  {\doibase 10.1103/PhysRevLett.120.057202} {\bibfield  {journal} {\bibinfo
  {journal} {Phys. Rev. Lett.}\ }\textbf {\bibinfo {volume} {120}},\ \bibinfo
  {pages} {057202} (\bibinfo {year} {2018})}\BibitemShut {NoStop}%
\bibitem [{\citenamefont {Yuan}\ \emph {et~al.}(2020)\citenamefont {Yuan},
  \citenamefont {Yan}, \citenamefont {Zheng}, \citenamefont {He}, \citenamefont
  {Xia},\ and\ \citenamefont {Yung}}]{PhysRevLett.124.053602}%
  \BibitemOpen
  \bibfield  {author} {\bibinfo {author} {\bibfnamefont {H.~Y.}\ \bibnamefont
  {Yuan}}, \bibinfo {author} {\bibfnamefont {P.}~\bibnamefont {Yan}}, \bibinfo
  {author} {\bibfnamefont {S.}~\bibnamefont {Zheng}}, \bibinfo {author}
  {\bibfnamefont {Q.~Y.}\ \bibnamefont {He}}, \bibinfo {author} {\bibfnamefont
  {K.}~\bibnamefont {Xia}}, \ and\ \bibinfo {author} {\bibfnamefont {M.-H.}\
  \bibnamefont {Yung}},\ }\bibfield  {title} {\enquote {\bibinfo {title}
  {Steady bell state generation via magnon-photon coupling},}\ }\href {\doibase
  10.1103/PhysRevLett.124.053602} {\bibfield  {journal} {\bibinfo  {journal}
  {Phys. Rev. Lett.}\ }\textbf {\bibinfo {volume} {124}},\ \bibinfo {pages}
  {053602} (\bibinfo {year} {2020})}\BibitemShut {NoStop}%
\bibitem [{\citenamefont {Huebl}\ \emph {et~al.}(2013)\citenamefont {Huebl},
  \citenamefont {Zollitsch}, \citenamefont {Lotze}, \citenamefont {Hocke},
  \citenamefont {Greifenstein}, \citenamefont {Marx}, \citenamefont {Gross},\
  and\ \citenamefont {Goennenwein}}]{PhysRevLett.111.127003}%
  \BibitemOpen
  \bibfield  {author} {\bibinfo {author} {\bibfnamefont {H.}~\bibnamefont
  {Huebl}}, \bibinfo {author} {\bibfnamefont {C.~W.}\ \bibnamefont
  {Zollitsch}}, \bibinfo {author} {\bibfnamefont {J.}~\bibnamefont {Lotze}},
  \bibinfo {author} {\bibfnamefont {F.}~\bibnamefont {Hocke}}, \bibinfo
  {author} {\bibfnamefont {M.}~\bibnamefont {Greifenstein}}, \bibinfo {author}
  {\bibfnamefont {A.}~\bibnamefont {Marx}}, \bibinfo {author} {\bibfnamefont
  {R.}~\bibnamefont {Gross}}, \ and\ \bibinfo {author} {\bibfnamefont
  {S.~T.~B.}\ \bibnamefont {Goennenwein}},\ }\bibfield  {title} {\enquote
  {\bibinfo {title} {High cooperativity in coupled microwave resonator
  ferrimagnetic insulator hybrids},}\ }\href {\doibase
  10.1103/PhysRevLett.111.127003} {\bibfield  {journal} {\bibinfo  {journal}
  {Phys. Rev. Lett.}\ }\textbf {\bibinfo {volume} {111}},\ \bibinfo {pages}
  {127003} (\bibinfo {year} {2013})}\BibitemShut {NoStop}%
\bibitem [{\citenamefont {Tabuchi}\ \emph {et~al.}(2014)\citenamefont
  {Tabuchi}, \citenamefont {Ishino}, \citenamefont {Ishikawa}, \citenamefont
  {Yamazaki}, \citenamefont {Usami},\ and\ \citenamefont
  {Nakamura}}]{PhysRevLett.113.083603}%
  \BibitemOpen
  \bibfield  {author} {\bibinfo {author} {\bibfnamefont {Y.}~\bibnamefont
  {Tabuchi}}, \bibinfo {author} {\bibfnamefont {S.}~\bibnamefont {Ishino}},
  \bibinfo {author} {\bibfnamefont {T.}~\bibnamefont {Ishikawa}}, \bibinfo
  {author} {\bibfnamefont {R.}~\bibnamefont {Yamazaki}}, \bibinfo {author}
  {\bibfnamefont {K.}~\bibnamefont {Usami}}, \ and\ \bibinfo {author}
  {\bibfnamefont {Y.}~\bibnamefont {Nakamura}},\ }\bibfield  {title} {\enquote
  {\bibinfo {title} {Hybridizing ferromagnetic magnons and microwave photons in
  the quantum limit},}\ }\href {\doibase 10.1103/PhysRevLett.113.083603}
  {\bibfield  {journal} {\bibinfo  {journal} {Phys. Rev. Lett.}\ }\textbf
  {\bibinfo {volume} {113}},\ \bibinfo {pages} {083603} (\bibinfo {year}
  {2014})}\BibitemShut {NoStop}%
\bibitem [{\citenamefont {Zhang}\ \emph {et~al.}(2014)\citenamefont {Zhang},
  \citenamefont {Zou}, \citenamefont {Jiang},\ and\ \citenamefont
  {Tang}}]{PhysRevLett.113.156401}%
  \BibitemOpen
  \bibfield  {author} {\bibinfo {author} {\bibfnamefont {X.}~\bibnamefont
  {Zhang}}, \bibinfo {author} {\bibfnamefont {C.-L.}\ \bibnamefont {Zou}},
  \bibinfo {author} {\bibfnamefont {L.}~\bibnamefont {Jiang}}, \ and\ \bibinfo
  {author} {\bibfnamefont {H.~X.}\ \bibnamefont {Tang}},\ }\bibfield  {title}
  {\enquote {\bibinfo {title} {Strongly coupled magnons and cavity microwave
  photons},}\ }\href {\doibase 10.1103/PhysRevLett.113.156401} {\bibfield
  {journal} {\bibinfo  {journal} {Phys. Rev. Lett.}\ }\textbf {\bibinfo
  {volume} {113}},\ \bibinfo {pages} {156401} (\bibinfo {year}
  {2014})}\BibitemShut {NoStop}%
\bibitem [{\citenamefont {Goryachev}\ \emph {et~al.}(2014)\citenamefont
  {Goryachev}, \citenamefont {Farr}, \citenamefont {Creedon}, \citenamefont
  {Fan}, \citenamefont {Kostylev},\ and\ \citenamefont
  {Tobar}}]{PhysRevApplied.2.054002}%
  \BibitemOpen
  \bibfield  {author} {\bibinfo {author} {\bibfnamefont {M.}~\bibnamefont
  {Goryachev}}, \bibinfo {author} {\bibfnamefont {W.~G.}\ \bibnamefont {Farr}},
  \bibinfo {author} {\bibfnamefont {D.~L.}\ \bibnamefont {Creedon}}, \bibinfo
  {author} {\bibfnamefont {Y.}~\bibnamefont {Fan}}, \bibinfo {author}
  {\bibfnamefont {M.}~\bibnamefont {Kostylev}}, \ and\ \bibinfo {author}
  {\bibfnamefont {M.~E.}\ \bibnamefont {Tobar}},\ }\bibfield  {title} {\enquote
  {\bibinfo {title} {High-cooperativity cavity qed with magnons at microwave
  frequencies},}\ }\href {\doibase 10.1103/PhysRevApplied.2.054002} {\bibfield
  {journal} {\bibinfo  {journal} {Phys. Rev. Applied}\ }\textbf {\bibinfo
  {volume} {2}},\ \bibinfo {pages} {054002} (\bibinfo {year}
  {2014})}\BibitemShut {NoStop}%
\bibitem [{\citenamefont {Zhang}\ \emph
  {et~al.}(2015{\natexlab{b}})\citenamefont {Zhang}, \citenamefont {Wang},
  \citenamefont {Li}, \citenamefont {Luo}, \citenamefont {Wu}, \citenamefont
  {Nori},\ and\ \citenamefont {You}}]{Zhang2015}%
  \BibitemOpen
  \bibfield  {author} {\bibinfo {author} {\bibfnamefont {D.}~\bibnamefont
  {Zhang}}, \bibinfo {author} {\bibfnamefont {X.-M.}\ \bibnamefont {Wang}},
  \bibinfo {author} {\bibfnamefont {T.-F.}\ \bibnamefont {Li}}, \bibinfo
  {author} {\bibfnamefont {X.-Q.}\ \bibnamefont {Luo}}, \bibinfo {author}
  {\bibfnamefont {W.}~\bibnamefont {Wu}}, \bibinfo {author} {\bibfnamefont
  {F.}~\bibnamefont {Nori}}, \ and\ \bibinfo {author} {\bibfnamefont {J.~Q.}\
  \bibnamefont {You}},\ }\bibfield  {title} {\enquote {\bibinfo {title} {Cavity
  quantum electrodynamics with ferromagnetic magnons in a small
  yttrium-iron-garnet sphere},}\ }\href {\doibase 10.1038/npjqi.2015.14}
  {\bibfield  {journal} {\bibinfo  {journal} {npj Quantum Information}\
  }\textbf {\bibinfo {volume} {1}},\ \bibinfo {pages} {15014} (\bibinfo {year}
  {2015}{\natexlab{b}})}\BibitemShut {NoStop}%
\bibitem [{\citenamefont {Tabuchi}\ \emph {et~al.}(2015)\citenamefont
  {Tabuchi}, \citenamefont {Ishino}, \citenamefont {Noguchi}, \citenamefont
  {Ishikawa}, \citenamefont {Yamazaki}, \citenamefont {Usami},\ and\
  \citenamefont {Nakamura}}]{Tabuchi405}%
  \BibitemOpen
  \bibfield  {author} {\bibinfo {author} {\bibfnamefont {Y.}~\bibnamefont
  {Tabuchi}}, \bibinfo {author} {\bibfnamefont {S.}~\bibnamefont {Ishino}},
  \bibinfo {author} {\bibfnamefont {A.}~\bibnamefont {Noguchi}}, \bibinfo
  {author} {\bibfnamefont {T.}~\bibnamefont {Ishikawa}}, \bibinfo {author}
  {\bibfnamefont {R.}~\bibnamefont {Yamazaki}}, \bibinfo {author}
  {\bibfnamefont {K.}~\bibnamefont {Usami}}, \ and\ \bibinfo {author}
  {\bibfnamefont {Y.}~\bibnamefont {Nakamura}},\ }\bibfield  {title} {\enquote
  {\bibinfo {title} {Coherent coupling between a ferromagnetic magnon and a
  superconducting qubit},}\ }\href {\doibase 10.1126/science.aaa3693}
  {\bibfield  {journal} {\bibinfo  {journal} {Science}\ }\textbf {\bibinfo
  {volume} {349}},\ \bibinfo {pages} {405--408} (\bibinfo {year} {2015})},\
  \Eprint
  {http://arxiv.org/abs/https://science.sciencemag.org/content/349/6246/405.full.pdf}
  {https://science.sciencemag.org/content/349/6246/405.full.pdf} \BibitemShut
  {NoStop}%
\bibitem [{\citenamefont {Wolz}\ \emph {et~al.}(2020)\citenamefont {Wolz},
  \citenamefont {Stehli}, \citenamefont {Schneider}, \citenamefont {Boventer},
  \citenamefont {Mac{\^e}do}, \citenamefont {Ustinov}, \citenamefont
  {Kl{\"a}ui},\ and\ \citenamefont {Weides}}]{Wolz2020}%
  \BibitemOpen
  \bibfield  {author} {\bibinfo {author} {\bibfnamefont {T.}~\bibnamefont
  {Wolz}}, \bibinfo {author} {\bibfnamefont {A.}~\bibnamefont {Stehli}},
  \bibinfo {author} {\bibfnamefont {A.}~\bibnamefont {Schneider}}, \bibinfo
  {author} {\bibfnamefont {I.}~\bibnamefont {Boventer}}, \bibinfo {author}
  {\bibfnamefont {R.}~\bibnamefont {Mac{\^e}do}}, \bibinfo {author}
  {\bibfnamefont {A.~V.}\ \bibnamefont {Ustinov}}, \bibinfo {author}
  {\bibfnamefont {M.}~\bibnamefont {Kl{\"a}ui}}, \ and\ \bibinfo {author}
  {\bibfnamefont {M.}~\bibnamefont {Weides}},\ }\bibfield  {title} {\enquote
  {\bibinfo {title} {Introducing coherent time control to cavity
  magnon-polariton modes},}\ }\href {\doibase 10.1038/s42005-019-0266-x}
  {\bibfield  {journal} {\bibinfo  {journal} {Communications Physics}\ }\textbf
  {\bibinfo {volume} {3}},\ \bibinfo {pages} {3} (\bibinfo {year}
  {2020})}\BibitemShut {NoStop}%
\bibitem [{\citenamefont {Roy}\ and\ \citenamefont
  {Devoret}(2016)}]{ROY2016740}%
  \BibitemOpen
  \bibfield  {author} {\bibinfo {author} {\bibfnamefont {A.}~\bibnamefont
  {Roy}}\ and\ \bibinfo {author} {\bibfnamefont {M.}~\bibnamefont {Devoret}},\
  }\bibfield  {title} {\enquote {\bibinfo {title} {Introduction to parametric
  amplification of quantum signals with josephson circuits},}\ }\href {\doibase
  https://doi.org/10.1016/j.crhy.2016.07.012} {\bibfield  {journal} {\bibinfo
  {journal} {Comptes Rendus Physique}\ }\textbf {\bibinfo {volume} {17}},\
  \bibinfo {pages} {740 -- 755} (\bibinfo {year} {2016})},\ \bibinfo {note}
  {quantum microwaves / Micro-ondes quantiques}\BibitemShut {NoStop}%
\bibitem [{\citenamefont {Bloom}(1957)}]{doi:10.1063/1.1722859}%
  \BibitemOpen
  \bibfield  {author} {\bibinfo {author} {\bibfnamefont {S.}~\bibnamefont
  {Bloom}},\ }\bibfield  {title} {\enquote {\bibinfo {title} {Effects of
  radiation damping on spin dynamics},}\ }\href {\doibase 10.1063/1.1722859}
  {\bibfield  {journal} {\bibinfo  {journal} {Journal of Applied Physics}\
  }\textbf {\bibinfo {volume} {28}},\ \bibinfo {pages} {800--805} (\bibinfo
  {year} {1957})}\BibitemShut {NoStop}%
\bibitem [{\citenamefont {Augustine}(2002)}]{AUGUSTINE2002111}%
  \BibitemOpen
  \bibfield  {author} {\bibinfo {author} {\bibfnamefont {M.}~\bibnamefont
  {Augustine}},\ }\bibfield  {title} {\enquote {\bibinfo {title} {Transient
  properties of radiation damping},}\ }\href {\doibase
  https://doi.org/10.1016/S0079-6565(01)00037-1} {\bibfield  {journal}
  {\bibinfo  {journal} {Progress in Nuclear Magnetic Resonance Spectroscopy}\
  }\textbf {\bibinfo {volume} {40}},\ \bibinfo {pages} {111 -- 150} (\bibinfo
  {year} {2002})}\BibitemShut {NoStop}%
\bibitem [{\citenamefont {Bloembergen}\ and\ \citenamefont
  {Pound}(1954)}]{PhysRev.95.8}%
  \BibitemOpen
  \bibfield  {author} {\bibinfo {author} {\bibfnamefont {N.}~\bibnamefont
  {Bloembergen}}\ and\ \bibinfo {author} {\bibfnamefont {R.~V.}\ \bibnamefont
  {Pound}},\ }\bibfield  {title} {\enquote {\bibinfo {title} {Radiation damping
  in magnetic resonance experiments},}\ }\href {\doibase 10.1103/PhysRev.95.8}
  {\bibfield  {journal} {\bibinfo  {journal} {Phys. Rev.}\ }\textbf {\bibinfo
  {volume} {95}},\ \bibinfo {pages} {8--12} (\bibinfo {year}
  {1954})}\BibitemShut {NoStop}%
\bibitem [{\citenamefont {Barbieri}\ \emph {et~al.}(2017)\citenamefont
  {Barbieri}, \citenamefont {Braggio}, \citenamefont {Carugno}, \citenamefont
  {Gallo}, \citenamefont {Lombardi}, \citenamefont {Ortolan}, \citenamefont
  {Pengo}, \citenamefont {Ruoso},\ and\ \citenamefont
  {Speake}}]{BARBIERI2017135}%
  \BibitemOpen
  \bibfield  {author} {\bibinfo {author} {\bibfnamefont {R.}~\bibnamefont
  {Barbieri}}, \bibinfo {author} {\bibfnamefont {C.}~\bibnamefont {Braggio}},
  \bibinfo {author} {\bibfnamefont {G.}~\bibnamefont {Carugno}}, \bibinfo
  {author} {\bibfnamefont {C.}~\bibnamefont {Gallo}}, \bibinfo {author}
  {\bibfnamefont {A.}~\bibnamefont {Lombardi}}, \bibinfo {author}
  {\bibfnamefont {A.}~\bibnamefont {Ortolan}}, \bibinfo {author} {\bibfnamefont
  {R.}~\bibnamefont {Pengo}}, \bibinfo {author} {\bibfnamefont
  {G.}~\bibnamefont {Ruoso}}, \ and\ \bibinfo {author} {\bibfnamefont
  {C.}~\bibnamefont {Speake}},\ }\bibfield  {title} {\enquote {\bibinfo {title}
  {Searching for galactic axions through magnetized media: The quax
  proposal},}\ }\href {\doibase https://doi.org/10.1016/j.dark.2017.01.003}
  {\bibfield  {journal} {\bibinfo  {journal} {Physics of the Dark Universe}\
  }\textbf {\bibinfo {volume} {15}},\ \bibinfo {pages} {135 -- 141} (\bibinfo
  {year} {2017})}\BibitemShut {NoStop}%
\bibitem [{\citenamefont {Barbieri}\ \emph {et~al.}(1989)\citenamefont
  {Barbieri}, \citenamefont {Cerdonio}, \citenamefont {Fiorentini},\ and\
  \citenamefont {Vitale}}]{BARBIERI1989357}%
  \BibitemOpen
  \bibfield  {author} {\bibinfo {author} {\bibfnamefont {R.}~\bibnamefont
  {Barbieri}}, \bibinfo {author} {\bibfnamefont {M.}~\bibnamefont {Cerdonio}},
  \bibinfo {author} {\bibfnamefont {G.}~\bibnamefont {Fiorentini}}, \ and\
  \bibinfo {author} {\bibfnamefont {S.}~\bibnamefont {Vitale}},\ }\bibfield
  {title} {\enquote {\bibinfo {title} {Axion to magnon conversion. a scheme for
  the detection of galactic axions},}\ }\href {\doibase
  https://doi.org/10.1016/0370-2693(89)91209-4} {\bibfield  {journal} {\bibinfo
   {journal} {Physics Letters B}\ }\textbf {\bibinfo {volume} {226}},\ \bibinfo
  {pages} {357 -- 360} (\bibinfo {year} {1989})}\BibitemShut {NoStop}%
\bibitem [{\citenamefont {{Crescini, N.}}\ \emph {et~al.}(2018)\citenamefont
  {{Crescini, N.}}, \citenamefont {{Alesini, D.}}, \citenamefont {{Braggio,
  C.}}, \citenamefont {{Carugno, G.}}, \citenamefont {{Di Gioacchino, D.}},
  \citenamefont {{Gallo, C. S.}}, \citenamefont {{Gambardella, U.}},
  \citenamefont {{Gatti, C.}}, \citenamefont {{Iannone, G.}}, \citenamefont
  {{Lamanna, G.}}, \citenamefont {{Ligi, C.}}, \citenamefont {{Lombardi, A.}},
  \citenamefont {{Ortolan, A.}}, \citenamefont {{Pagano, S.}}, \citenamefont
  {{Pengo, R.}}, \citenamefont {{Ruoso, G.}}, \citenamefont {{Speake, C. C.}},\
  and\ \citenamefont {{Taffarello, L.}}}]{quaxepjc}%
  \BibitemOpen
  \bibfield  {author} {\bibinfo {author} {\bibnamefont {{Crescini, N.}}},
  \bibinfo {author} {\bibnamefont {{Alesini, D.}}}, \bibinfo {author}
  {\bibnamefont {{Braggio, C.}}}, \bibinfo {author} {\bibnamefont {{Carugno,
  G.}}}, \bibinfo {author} {\bibnamefont {{Di Gioacchino, D.}}}, \bibinfo
  {author} {\bibnamefont {{Gallo, C. S.}}}, \bibinfo {author} {\bibnamefont
  {{Gambardella, U.}}}, \bibinfo {author} {\bibnamefont {{Gatti, C.}}},
  \bibinfo {author} {\bibnamefont {{Iannone, G.}}}, \bibinfo {author}
  {\bibnamefont {{Lamanna, G.}}}, \bibinfo {author} {\bibnamefont {{Ligi,
  C.}}}, \bibinfo {author} {\bibnamefont {{Lombardi, A.}}}, \bibinfo {author}
  {\bibnamefont {{Ortolan, A.}}}, \bibinfo {author} {\bibnamefont {{Pagano,
  S.}}}, \bibinfo {author} {\bibnamefont {{Pengo, R.}}}, \bibinfo {author}
  {\bibnamefont {{Ruoso, G.}}}, \bibinfo {author} {\bibnamefont {{Speake, C.
  C.}}}, \ and\ \bibinfo {author} {\bibnamefont {{Taffarello, L.}}},\
  }\bibfield  {title} {\enquote {\bibinfo {title} {Operation of a ferromagnetic
  axion haloscope at $m_a=58\,\mu\mathrm{eV}$},}\ }\href {\doibase
  10.1140/epjc/s10052-018-6163-8} {\bibfield  {journal} {\bibinfo  {journal}
  {Eur. Phys. J. C}\ }\textbf {\bibinfo {volume} {78}},\ \bibinfo {pages} {703}
  (\bibinfo {year} {2018})}\BibitemShut {NoStop}%
\bibitem [{\citenamefont {Crescini}\ \emph
  {et~al.}(2020{\natexlab{a}})\citenamefont {Crescini}, \citenamefont
  {Alesini}, \citenamefont {Braggio}, \citenamefont {Carugno}, \citenamefont
  {D'Agostino}, \citenamefont {Di~Gioacchino}, \citenamefont {Falferi},
  \citenamefont {Gambardella}, \citenamefont {Gatti}, \citenamefont {Iannone},
  \citenamefont {Ligi}, \citenamefont {Lombardi}, \citenamefont {Ortolan},
  \citenamefont {Pengo}, \citenamefont {Ruoso},\ and\ \citenamefont
  {Taffarello}}]{PhysRevLett.124.171801}%
  \BibitemOpen
  \bibfield  {author} {\bibinfo {author} {\bibfnamefont {N.}~\bibnamefont
  {Crescini}}, \bibinfo {author} {\bibfnamefont {D.}~\bibnamefont {Alesini}},
  \bibinfo {author} {\bibfnamefont {C.}~\bibnamefont {Braggio}}, \bibinfo
  {author} {\bibfnamefont {G.}~\bibnamefont {Carugno}}, \bibinfo {author}
  {\bibfnamefont {D.}~\bibnamefont {D'Agostino}}, \bibinfo {author}
  {\bibfnamefont {D.}~\bibnamefont {Di~Gioacchino}}, \bibinfo {author}
  {\bibfnamefont {P.}~\bibnamefont {Falferi}}, \bibinfo {author} {\bibfnamefont
  {U.}~\bibnamefont {Gambardella}}, \bibinfo {author} {\bibfnamefont
  {C.}~\bibnamefont {Gatti}}, \bibinfo {author} {\bibfnamefont
  {G.}~\bibnamefont {Iannone}}, \bibinfo {author} {\bibfnamefont
  {C.}~\bibnamefont {Ligi}}, \bibinfo {author} {\bibfnamefont {A.}~\bibnamefont
  {Lombardi}}, \bibinfo {author} {\bibfnamefont {A.}~\bibnamefont {Ortolan}},
  \bibinfo {author} {\bibfnamefont {R.}~\bibnamefont {Pengo}}, \bibinfo
  {author} {\bibfnamefont {G.}~\bibnamefont {Ruoso}}, \ and\ \bibinfo {author}
  {\bibfnamefont {L.}~\bibnamefont {Taffarello}} (\bibinfo {collaboration}
  {QUAX Collaboration}),\ }\bibfield  {title} {\enquote {\bibinfo {title}
  {Axion search with a quantum-limited ferromagnetic haloscope},}\ }\href
  {\doibase 10.1103/PhysRevLett.124.171801} {\bibfield  {journal} {\bibinfo
  {journal} {Phys. Rev. Lett.}\ }\textbf {\bibinfo {volume} {124}},\ \bibinfo
  {pages} {171801} (\bibinfo {year} {2020}{\natexlab{a}})}\BibitemShut
  {NoStop}%
\bibitem [{\citenamefont {Crescini}\ \emph
  {et~al.}(2020{\natexlab{b}})\citenamefont {Crescini}, \citenamefont
  {Braggio}, \citenamefont {Carugno}, \citenamefont {Ortolan},\ and\
  \citenamefont {Ruoso}}]{crescini2020coherent}%
  \BibitemOpen
  \bibfield  {author} {\bibinfo {author} {\bibfnamefont {N.}~\bibnamefont
  {Crescini}}, \bibinfo {author} {\bibfnamefont {C.}~\bibnamefont {Braggio}},
  \bibinfo {author} {\bibfnamefont {G.}~\bibnamefont {Carugno}}, \bibinfo
  {author} {\bibfnamefont {A.}~\bibnamefont {Ortolan}}, \ and\ \bibinfo
  {author} {\bibfnamefont {G.}~\bibnamefont {Ruoso}},\ }\href
  {http://arxiv.org/abs/2007.08908} {\enquote {\bibinfo {title} {Coherent
  coupling between multiple ferrimagnetic spheres and a microwave cavity in the
  quantum-limit},}\ } (\bibinfo {year} {2020}{\natexlab{b}}),\ \Eprint
  {http://arxiv.org/abs/2007.08908} {arXiv:2007.08908 [quant-ph]} \BibitemShut
  {NoStop}%
\bibitem [{\citenamefont {Macêdo}\ \emph {et~al.}(2020)\citenamefont
  {Macêdo}, \citenamefont {Holland}, \citenamefont {Baity}, \citenamefont
  {Livesey}, \citenamefont {Stamps}, \citenamefont {Weides},\ and\
  \citenamefont {Bozhko}}]{paperdelreferee}%
  \BibitemOpen
  \bibfield  {author} {\bibinfo {author} {\bibfnamefont {R.}~\bibnamefont
  {Macêdo}}, \bibinfo {author} {\bibfnamefont {R.~C.}\ \bibnamefont
  {Holland}}, \bibinfo {author} {\bibfnamefont {P.~G.}\ \bibnamefont {Baity}},
  \bibinfo {author} {\bibfnamefont {K.~L.}\ \bibnamefont {Livesey}}, \bibinfo
  {author} {\bibfnamefont {R.~L.}\ \bibnamefont {Stamps}}, \bibinfo {author}
  {\bibfnamefont {M.~P.}\ \bibnamefont {Weides}}, \ and\ \bibinfo {author}
  {\bibfnamefont {D.~A.}\ \bibnamefont {Bozhko}},\ }\href
  {https://arxiv.org/abs/2007.11483} {\enquote {\bibinfo {title} {An
  electromagnetic approach to cavity spintronics},}\ } (\bibinfo {year}
  {2020}),\ \Eprint {http://arxiv.org/abs/arXiv:2007.11483}
  {arXiv:arXiv:2007.11483 [quant-ph]} \BibitemShut {NoStop}%
\bibitem [{\citenamefont {Roch}\ \emph {et~al.}(2012)\citenamefont {Roch},
  \citenamefont {Flurin}, \citenamefont {Nguyen}, \citenamefont {Morfin},
  \citenamefont {Campagne-Ibarcq}, \citenamefont {Devoret},\ and\ \citenamefont
  {Huard}}]{PhysRevLett.108.147701}%
  \BibitemOpen
  \bibfield  {author} {\bibinfo {author} {\bibfnamefont {N.}~\bibnamefont
  {Roch}}, \bibinfo {author} {\bibfnamefont {E.}~\bibnamefont {Flurin}},
  \bibinfo {author} {\bibfnamefont {F.}~\bibnamefont {Nguyen}}, \bibinfo
  {author} {\bibfnamefont {P.}~\bibnamefont {Morfin}}, \bibinfo {author}
  {\bibfnamefont {P.}~\bibnamefont {Campagne-Ibarcq}}, \bibinfo {author}
  {\bibfnamefont {M.~H.}\ \bibnamefont {Devoret}}, \ and\ \bibinfo {author}
  {\bibfnamefont {B.}~\bibnamefont {Huard}},\ }\bibfield  {title} {\enquote
  {\bibinfo {title} {Widely tunable, nondegenerate three-wave mixing microwave
  device operating near the quantum limit},}\ }\href {\doibase
  10.1103/PhysRevLett.108.147701} {\bibfield  {journal} {\bibinfo  {journal}
  {Phys. Rev. Lett.}\ }\textbf {\bibinfo {volume} {108}},\ \bibinfo {pages}
  {147701} (\bibinfo {year} {2012})}\BibitemShut {NoStop}%
\bibitem [{\citenamefont {Jaklevic}\ \emph {et~al.}(1964)\citenamefont
  {Jaklevic}, \citenamefont {Lambe}, \citenamefont {Silver},\ and\
  \citenamefont {Mercereau}}]{PhysRevLett.12.159}%
  \BibitemOpen
  \bibfield  {author} {\bibinfo {author} {\bibfnamefont {R.~C.}\ \bibnamefont
  {Jaklevic}}, \bibinfo {author} {\bibfnamefont {J.}~\bibnamefont {Lambe}},
  \bibinfo {author} {\bibfnamefont {A.~H.}\ \bibnamefont {Silver}}, \ and\
  \bibinfo {author} {\bibfnamefont {J.~E.}\ \bibnamefont {Mercereau}},\
  }\bibfield  {title} {\enquote {\bibinfo {title} {Quantum interference effects
  in josephson tunneling},}\ }\href {\doibase 10.1103/PhysRevLett.12.159}
  {\bibfield  {journal} {\bibinfo  {journal} {Phys. Rev. Lett.}\ }\textbf
  {\bibinfo {volume} {12}},\ \bibinfo {pages} {159--160} (\bibinfo {year}
  {1964})}\BibitemShut {NoStop}%
\bibitem [{\citenamefont {Erné}, \citenamefont {Hahlbohm},\ and\ \citenamefont
  {Lübbig}(1976)}]{doi:10.1063/1.322574}%
  \BibitemOpen
  \bibfield  {author} {\bibinfo {author} {\bibfnamefont {S.~N.}\ \bibnamefont
  {Erné}}, \bibinfo {author} {\bibfnamefont {H.}~\bibnamefont {Hahlbohm}}, \
  and\ \bibinfo {author} {\bibfnamefont {H.}~\bibnamefont {Lübbig}},\
  }\bibfield  {title} {\enquote {\bibinfo {title} {Theory of rf‐biased
  superconducting quantum interference device for nonhysteretic regime},}\
  }\href {\doibase 10.1063/1.322574} {\bibfield  {journal} {\bibinfo  {journal}
  {Journal of Applied Physics}\ }\textbf {\bibinfo {volume} {47}},\ \bibinfo
  {pages} {5440--5442} (\bibinfo {year} {1976})},\ \Eprint
  {http://arxiv.org/abs/https://doi.org/10.1063/1.322574}
  {https://doi.org/10.1063/1.322574} \BibitemShut {NoStop}%
\bibitem [{\citenamefont {Aprili}(2006)}]{Aprili2006}%
  \BibitemOpen
  \bibfield  {author} {\bibinfo {author} {\bibfnamefont {M.}~\bibnamefont
  {Aprili}},\ }\bibfield  {title} {\enquote {\bibinfo {title} {The nanosquid
  makes its debut},}\ }\href {\doibase 10.1038/nnano.2006.78} {\bibfield
  {journal} {\bibinfo  {journal} {Nature Nanotechnology}\ }\textbf {\bibinfo
  {volume} {1}},\ \bibinfo {pages} {15--16} (\bibinfo {year}
  {2006})}\BibitemShut {NoStop}%
\bibitem [{\citenamefont {{Kleiner}}\ \emph {et~al.}(2004)\citenamefont
  {{Kleiner}}, \citenamefont {{Koelle}}, \citenamefont {{Ludwig}},\ and\
  \citenamefont {{Clarke}}}]{1335547}%
  \BibitemOpen
  \bibfield  {author} {\bibinfo {author} {\bibfnamefont {R.}~\bibnamefont
  {{Kleiner}}}, \bibinfo {author} {\bibfnamefont {D.}~\bibnamefont {{Koelle}}},
  \bibinfo {author} {\bibfnamefont {F.}~\bibnamefont {{Ludwig}}}, \ and\
  \bibinfo {author} {\bibfnamefont {J.}~\bibnamefont {{Clarke}}},\ }\bibfield
  {title} {\enquote {\bibinfo {title} {Superconducting quantum interference
  devices: State of the art and applications},}\ }\href {\doibase
  10.1109/JPROC.2004.833655} {\bibfield  {journal} {\bibinfo  {journal}
  {Proceedings of the IEEE}\ }\textbf {\bibinfo {volume} {92}},\ \bibinfo
  {pages} {1534--1548} (\bibinfo {year} {2004})}\BibitemShut {NoStop}%
\bibitem [{\citenamefont {John}\ and\ \citenamefont
  {Alex~I.}(2005)}]{doi:10.1002/3527603646}%
  \BibitemOpen
  \bibfield  {author} {\bibinfo {author} {\bibfnamefont {C.}~\bibnamefont
  {John}}\ and\ \bibinfo {author} {\bibfnamefont {B.}~\bibnamefont {Alex~I.}},\
  }\href {\doibase 10.1002/3527603646.fmatter} {\emph {\bibinfo {title} {The
  SQUID Handbook}}}\ (\bibinfo  {publisher} {John Wiley \& Sons, Ltd},\
  \bibinfo {year} {2005})\ \Eprint
  {http://arxiv.org/abs/https://onlinelibrary.wiley.com/doi/pdf/10.1002/3527603646.fmatter}
  {https://onlinelibrary.wiley.com/doi/pdf/10.1002/3527603646.fmatter}
  \BibitemShut {NoStop}%
\bibitem [{\citenamefont {Kominis}\ \emph {et~al.}(2003)\citenamefont
  {Kominis}, \citenamefont {Kornack}, \citenamefont {Allred},\ and\
  \citenamefont {Romalis}}]{Kominis2003}%
  \BibitemOpen
  \bibfield  {author} {\bibinfo {author} {\bibfnamefont {I.~K.}\ \bibnamefont
  {Kominis}}, \bibinfo {author} {\bibfnamefont {T.~W.}\ \bibnamefont
  {Kornack}}, \bibinfo {author} {\bibfnamefont {J.~C.}\ \bibnamefont {Allred}},
  \ and\ \bibinfo {author} {\bibfnamefont {M.~V.}\ \bibnamefont {Romalis}},\
  }\bibfield  {title} {\enquote {\bibinfo {title} {A subfemtotesla multichannel
  atomic magnetometer},}\ }\href {\doibase 10.1038/nature01484} {\bibfield
  {journal} {\bibinfo  {journal} {Nature}\ }\textbf {\bibinfo {volume} {422}},\
  \bibinfo {pages} {596--599} (\bibinfo {year} {2003})}\BibitemShut {NoStop}%
\bibitem [{\citenamefont {Savukov}\ \emph {et~al.}(2005)\citenamefont
  {Savukov}, \citenamefont {Seltzer}, \citenamefont {Romalis},\ and\
  \citenamefont {Sauer}}]{PhysRevLett.95.063004}%
  \BibitemOpen
  \bibfield  {author} {\bibinfo {author} {\bibfnamefont {I.~M.}\ \bibnamefont
  {Savukov}}, \bibinfo {author} {\bibfnamefont {S.~J.}\ \bibnamefont
  {Seltzer}}, \bibinfo {author} {\bibfnamefont {M.~V.}\ \bibnamefont
  {Romalis}}, \ and\ \bibinfo {author} {\bibfnamefont {K.~L.}\ \bibnamefont
  {Sauer}},\ }\bibfield  {title} {\enquote {\bibinfo {title} {Tunable atomic
  magnetometer for detection of radio-frequency magnetic fields},}\ }\href
  {\doibase 10.1103/PhysRevLett.95.063004} {\bibfield  {journal} {\bibinfo
  {journal} {Phys. Rev. Lett.}\ }\textbf {\bibinfo {volume} {95}},\ \bibinfo
  {pages} {063004} (\bibinfo {year} {2005})}\BibitemShut {NoStop}%
\bibitem [{\citenamefont {Budker}\ and\ \citenamefont
  {Romalis}(2007)}]{Budker2007}%
  \BibitemOpen
  \bibfield  {author} {\bibinfo {author} {\bibfnamefont {D.}~\bibnamefont
  {Budker}}\ and\ \bibinfo {author} {\bibfnamefont {M.}~\bibnamefont
  {Romalis}},\ }\bibfield  {title} {\enquote {\bibinfo {title} {Optical
  magnetometry},}\ }\href {\doibase 10.1038/nphys566} {\bibfield  {journal}
  {\bibinfo  {journal} {Nature Physics}\ }\textbf {\bibinfo {volume} {3}},\
  \bibinfo {pages} {227--234} (\bibinfo {year} {2007})}\BibitemShut {NoStop}%
\bibitem [{\citenamefont {Crescini}, \citenamefont {Ruoso},\ and\ \citenamefont
  {Carugno}(2020)}]{inprepmag}%
  \BibitemOpen
  \bibfield  {author} {\bibinfo {author} {\bibfnamefont {N.}~\bibnamefont
  {Crescini}}, \bibinfo {author} {\bibfnamefont {G.}~\bibnamefont {Ruoso}}, \
  and\ \bibinfo {author} {\bibfnamefont {G.}~\bibnamefont {Carugno}},\
  }\href@noop {} {\enquote {\bibinfo {title} {In preparation},}\ } (\bibinfo
  {year} {2020})\BibitemShut {NoStop}%
\bibitem [{\citenamefont {Kempf}\ \emph {et~al.}(2017)\citenamefont {Kempf},
  \citenamefont {Wegner}, \citenamefont {Fleischmann}, \citenamefont
  {Gastaldo}, \citenamefont {Herrmann}, \citenamefont {Papst}, \citenamefont
  {Richter},\ and\ \citenamefont {Enss}}]{doi:10.1063/1.4973872}%
  \BibitemOpen
  \bibfield  {author} {\bibinfo {author} {\bibfnamefont {S.}~\bibnamefont
  {Kempf}}, \bibinfo {author} {\bibfnamefont {M.}~\bibnamefont {Wegner}},
  \bibinfo {author} {\bibfnamefont {A.}~\bibnamefont {Fleischmann}}, \bibinfo
  {author} {\bibfnamefont {L.}~\bibnamefont {Gastaldo}}, \bibinfo {author}
  {\bibfnamefont {F.}~\bibnamefont {Herrmann}}, \bibinfo {author}
  {\bibfnamefont {M.}~\bibnamefont {Papst}}, \bibinfo {author} {\bibfnamefont
  {D.}~\bibnamefont {Richter}}, \ and\ \bibinfo {author} {\bibfnamefont
  {C.}~\bibnamefont {Enss}},\ }\bibfield  {title} {\enquote {\bibinfo {title}
  {Demonstration of a scalable frequency-domain readout of metallic magnetic
  calorimeters by means of a microwave squid multiplexer},}\ }\href {\doibase
  10.1063/1.4973872} {\bibfield  {journal} {\bibinfo  {journal} {AIP Advances}\
  }\textbf {\bibinfo {volume} {7}},\ \bibinfo {pages} {015007} (\bibinfo {year}
  {2017})},\ \Eprint {http://arxiv.org/abs/https://doi.org/10.1063/1.4973872}
  {https://doi.org/10.1063/1.4973872} \BibitemShut {NoStop}%
\bibitem [{\citenamefont {Irwin}\ and\ \citenamefont
  {Lehnert}(2004)}]{doi:10.1063/1.1791733}%
  \BibitemOpen
  \bibfield  {author} {\bibinfo {author} {\bibfnamefont {K.~D.}\ \bibnamefont
  {Irwin}}\ and\ \bibinfo {author} {\bibfnamefont {K.~W.}\ \bibnamefont
  {Lehnert}},\ }\bibfield  {title} {\enquote {\bibinfo {title} {Microwave squid
  multiplexer},}\ }\href {\doibase 10.1063/1.1791733} {\bibfield  {journal}
  {\bibinfo  {journal} {Applied Physics Letters}\ }\textbf {\bibinfo {volume}
  {85}},\ \bibinfo {pages} {2107--2109} (\bibinfo {year} {2004})},\ \Eprint
  {http://arxiv.org/abs/https://doi.org/10.1063/1.1791733}
  {https://doi.org/10.1063/1.1791733} \BibitemShut {NoStop}%
\bibitem [{\citenamefont {Mates}\ \emph {et~al.}(2008)\citenamefont {Mates},
  \citenamefont {Hilton}, \citenamefont {Irwin}, \citenamefont {Vale},\ and\
  \citenamefont {Lehnert}}]{doi:10.1063/1.2803852}%
  \BibitemOpen
  \bibfield  {author} {\bibinfo {author} {\bibfnamefont {J.~A.~B.}\
  \bibnamefont {Mates}}, \bibinfo {author} {\bibfnamefont {G.~C.}\ \bibnamefont
  {Hilton}}, \bibinfo {author} {\bibfnamefont {K.~D.}\ \bibnamefont {Irwin}},
  \bibinfo {author} {\bibfnamefont {L.~R.}\ \bibnamefont {Vale}}, \ and\
  \bibinfo {author} {\bibfnamefont {K.~W.}\ \bibnamefont {Lehnert}},\
  }\bibfield  {title} {\enquote {\bibinfo {title} {Demonstration of a
  multiplexer of dissipationless superconducting quantum interference
  devices},}\ }\href {\doibase 10.1063/1.2803852} {\bibfield  {journal}
  {\bibinfo  {journal} {Applied Physics Letters}\ }\textbf {\bibinfo {volume}
  {92}},\ \bibinfo {pages} {023514} (\bibinfo {year} {2008})},\ \Eprint
  {http://arxiv.org/abs/https://doi.org/10.1063/1.2803852}
  {https://doi.org/10.1063/1.2803852} \BibitemShut {NoStop}%
\bibitem [{\citenamefont {Lamoreaux}\ \emph {et~al.}(2013)\citenamefont
  {Lamoreaux}, \citenamefont {van Bibber}, \citenamefont {Lehnert},\ and\
  \citenamefont {Carosi}}]{PhysRevD.88.035020}%
  \BibitemOpen
  \bibfield  {author} {\bibinfo {author} {\bibfnamefont {S.~K.}\ \bibnamefont
  {Lamoreaux}}, \bibinfo {author} {\bibfnamefont {K.~A.}\ \bibnamefont {van
  Bibber}}, \bibinfo {author} {\bibfnamefont {K.~W.}\ \bibnamefont {Lehnert}},
  \ and\ \bibinfo {author} {\bibfnamefont {G.}~\bibnamefont {Carosi}},\
  }\bibfield  {title} {\enquote {\bibinfo {title} {Analysis of single-photon
  and linear amplifier detectors for microwave cavity dark matter axion
  searches},}\ }\href {\doibase 10.1103/PhysRevD.88.035020} {\bibfield
  {journal} {\bibinfo  {journal} {Phys. Rev. D}\ }\textbf {\bibinfo {volume}
  {88}},\ \bibinfo {pages} {035020} (\bibinfo {year} {2013})}\BibitemShut
  {NoStop}%
\bibitem [{\citenamefont {Cullen}(1958)}]{CULLEN1958}%
  \BibitemOpen
  \bibfield  {author} {\bibinfo {author} {\bibfnamefont {A.~L.}\ \bibnamefont
  {Cullen}},\ }\bibfield  {title} {\enquote {\bibinfo {title} {A
  travelling-wave parametric amplifier},}\ }\href {\doibase 10.1038/181332a0}
  {\bibfield  {journal} {\bibinfo  {journal} {Nature}\ }\textbf {\bibinfo
  {volume} {181}},\ \bibinfo {pages} {332--332} (\bibinfo {year}
  {1958})}\BibitemShut {NoStop}%
\bibitem [{\citenamefont {Macklin}\ \emph {et~al.}(2015)\citenamefont
  {Macklin}, \citenamefont {O{\textquoteright}Brien}, \citenamefont {Hover},
  \citenamefont {Schwartz}, \citenamefont {Bolkhovsky}, \citenamefont {Zhang},
  \citenamefont {Oliver},\ and\ \citenamefont {Siddiqi}}]{Macklin307}%
  \BibitemOpen
  \bibfield  {author} {\bibinfo {author} {\bibfnamefont {C.}~\bibnamefont
  {Macklin}}, \bibinfo {author} {\bibfnamefont {K.}~\bibnamefont
  {O{\textquoteright}Brien}}, \bibinfo {author} {\bibfnamefont
  {D.}~\bibnamefont {Hover}}, \bibinfo {author} {\bibfnamefont {M.~E.}\
  \bibnamefont {Schwartz}}, \bibinfo {author} {\bibfnamefont {V.}~\bibnamefont
  {Bolkhovsky}}, \bibinfo {author} {\bibfnamefont {X.}~\bibnamefont {Zhang}},
  \bibinfo {author} {\bibfnamefont {W.~D.}\ \bibnamefont {Oliver}}, \ and\
  \bibinfo {author} {\bibfnamefont {I.}~\bibnamefont {Siddiqi}},\ }\bibfield
  {title} {\enquote {\bibinfo {title} {A near{\textendash}quantum-limited
  josephson traveling-wave parametric amplifier},}\ }\href {\doibase
  10.1126/science.aaa8525} {\bibfield  {journal} {\bibinfo  {journal}
  {Science}\ }\textbf {\bibinfo {volume} {350}},\ \bibinfo {pages} {307--310}
  (\bibinfo {year} {2015})},\ \Eprint
  {http://arxiv.org/abs/https://science.sciencemag.org/content/350/6258/307.full.pdf}
  {https://science.sciencemag.org/content/350/6258/307.full.pdf} \BibitemShut
  {NoStop}%
\bibitem [{\citenamefont {Planat}\ \emph {et~al.}(2020)\citenamefont {Planat},
  \citenamefont {Ranadive}, \citenamefont {Dassonneville}, \citenamefont
  {Puertas~Mart\'{\i}nez}, \citenamefont {L\'eger}, \citenamefont {Naud},
  \citenamefont {Buisson}, \citenamefont {Hasch-Guichard}, \citenamefont
  {Basko},\ and\ \citenamefont {Roch}}]{PhysRevX.10.021021}%
  \BibitemOpen
  \bibfield  {author} {\bibinfo {author} {\bibfnamefont {L.}~\bibnamefont
  {Planat}}, \bibinfo {author} {\bibfnamefont {A.}~\bibnamefont {Ranadive}},
  \bibinfo {author} {\bibfnamefont {R.}~\bibnamefont {Dassonneville}}, \bibinfo
  {author} {\bibfnamefont {J.}~\bibnamefont {Puertas~Mart\'{\i}nez}}, \bibinfo
  {author} {\bibfnamefont {S.}~\bibnamefont {L\'eger}}, \bibinfo {author}
  {\bibfnamefont {C.}~\bibnamefont {Naud}}, \bibinfo {author} {\bibfnamefont
  {O.}~\bibnamefont {Buisson}}, \bibinfo {author} {\bibfnamefont
  {W.}~\bibnamefont {Hasch-Guichard}}, \bibinfo {author} {\bibfnamefont
  {D.~M.}\ \bibnamefont {Basko}}, \ and\ \bibinfo {author} {\bibfnamefont
  {N.}~\bibnamefont {Roch}},\ }\bibfield  {title} {\enquote {\bibinfo {title}
  {Photonic-crystal josephson traveling-wave parametric amplifier},}\ }\href
  {\doibase 10.1103/PhysRevX.10.021021} {\bibfield  {journal} {\bibinfo
  {journal} {Phys. Rev. X}\ }\textbf {\bibinfo {volume} {10}},\ \bibinfo
  {pages} {021021} (\bibinfo {year} {2020})}\BibitemShut {NoStop}%
\bibitem [{\citenamefont {Crescini}\ \emph
  {et~al.}(2020{\natexlab{c}})\citenamefont {Crescini}, \citenamefont
  {Braggio}, \citenamefont {Carugno}, \citenamefont {Di~Vora}, \citenamefont
  {Ortolan},\ and\ \citenamefont {Ruoso}}]{crescini2020magnon}%
  \BibitemOpen
  \bibfield  {author} {\bibinfo {author} {\bibfnamefont {N.}~\bibnamefont
  {Crescini}}, \bibinfo {author} {\bibfnamefont {C.}~\bibnamefont {Braggio}},
  \bibinfo {author} {\bibfnamefont {G.}~\bibnamefont {Carugno}}, \bibinfo
  {author} {\bibfnamefont {R.}~\bibnamefont {Di~Vora}}, \bibinfo {author}
  {\bibfnamefont {A.}~\bibnamefont {Ortolan}}, \ and\ \bibinfo {author}
  {\bibfnamefont {G.}~\bibnamefont {Ruoso}},\ }\bibfield  {title} {\enquote
  {\bibinfo {title} {Magnon-driven dynamics of a hybrid system excited with
  ultrafast optical pulses},}\ }\href {\doibase 10.1038/s42005-020-00435-w}
  {\bibfield  {journal} {\bibinfo  {journal} {Communications Physics}\ }\textbf
  {\bibinfo {volume} {3}},\ \bibinfo {pages} {164} (\bibinfo {year}
  {2020}{\natexlab{c}})}\BibitemShut {NoStop}%
\end{thebibliography}%

\end{document}